\documentclass[11pt]{article}

\def\sgn{\;{\rm sgn}\;}

\newcommand{\dd}[2]{\frac {\partial #1}{\partial #2}}
\newcommand{\pdr}{\partial}

\newcommand{\beq}{\begin{eqnarray}}
\newcommand{\eeq}{\end{eqnarray}}

\def\bra{\langle}   \def\ket{\rangle}

\newcommand{\half}{\frac{1}{2}}
\newcommand{\eps}{\epsilon}
\newcommand{\veps}{\varepsilon}
\newcommand{\om}{\omega}
\newcommand{\Om}{\Omega}

\newcommand{\ov}[1]{\frac{1}{#1}}
\newcommand{\fr}[2]{\frac{#1}{#2}}

\newcommand{\La}{\Lambda}

\newcommand{\tht}{\theta}
\newcommand{\Tht}{\Theta}
\newcommand{\sig}{\sigma}

\newcommand{\grad}{\nabla}
\newcommand{\g}{{\cal G}}
\newcommand{\gd}{{\cal G}^*}

\newcommand{\sdiff}{{\rm SDiff}(M,\mu)}
\newcommand{\diff}{{\rm Diff}(M)}
\newcommand{\svect}{{\rm SVect}(M,\mu)}
\newcommand{\A}{\tilde A}
\newcommand{\vect}{{\rm Vect}(M)}

\usepackage{hyperref}

\begin{document}
\begin{titlepage}

\title{\normalsize \hfill ITP-UU-05/28  \\ \hfill SPIN-05/22
\\ \hfill {\tt hep-th/0507283}\\ \vskip 0mm \Large\bf
$2+1$ Abelian `Gauge Theory' Inspired by Ideal Hydrodynamics}

\author{Govind S. Krishnaswami}
\date{\normalsize Institute for Theoretical Physics \& Spinoza Institute \\
Utrecht University, Postbus 80.195, 3508 TD, Utrecht, The
Netherlands
\smallskip \\ e-mail: {\tt g.s.krishnaswami@phys.uu.nl} \\
\medskip Journal Reference: Int.J.Mod.Phys.A21 (2006) 3771-3808, Received 11 October, 2005}

\maketitle

\begin{quotation} \noindent {\large\bf Abstract } \medskip \\

We study a possibly integrable model of abelian gauge fields on a
two-dimensional surface $M$, with volume form $\mu$. It has the same
phase space as ideal hydrodynamics, a coadjoint orbit of the
volume-preserving diffeomorphism group of $M$. Gauge field Poisson
brackets differ from the Heisenberg algebra, but are reminiscent of
Yang-Mills theory on a null surface. Enstrophy invariants are
Casimirs of the Poisson algebra of gauge invariant observables. Some
symplectic leaves of the Poisson manifold are identified. The
Hamiltonian is a magnetic energy, similar to that of
electrodynamics, and depends on a metric whose volume element is not
a multiple of $\mu$. The magnetic field evolves by a quadratically
non-linear `Euler' equation, which may also be regarded as
describing geodesic flow on $\sdiff$. Static solutions are obtained.
For uniform $\mu$, an infinite sequence of local conserved charges
beginning with the hamiltonian are found. The charges are shown to
be in involution, suggesting integrability. Besides being a theory
of a novel kind of ideal flow, this is a toy-model for Yang-Mills
theory and matrix field theories, whose gauge-invariant phase space
is conjectured to be a coadjoint orbit of the diffeomorphism group
of a non-commutative space.

\end{quotation}

\vfill \flushleft




\thispagestyle{empty}

\end{titlepage}

\eject

\section{Introduction and Summary}
\label{s-intro}

The classical theory of gauge fields\footnote{After gauge-fixing,
there will be a single propagating field degree of freedom.}
$A_1(x^1, x^2, t)$ and $A_2(x^1, x^2, t)$ we study in this paper,
may be summarized in four equations. The hamiltonian is a
gauge-invariant magnetic energy,
    \beq
    H = \int \bigg(\fr{B}{\rho} \bigg)^2 ~\sig ~\rho~ d^2x
    \eeq
where $B = \pdr_1 A_2 - \pdr_2 A_1$ is the magnetic field. $\rho$ is
a given volume element and $g_{ij}$ a fixed metric on a two
dimensional surface $M$ such that $\sigma = \rho^2/g$ is not a
constant ($g = \det{g_{ij}}$). The Poisson bracket between gauge
fields is a Lie algebra (independent of the metric)
    \beq
        \{A_i(x),A_j(y) \} = \delta^2(x-y) \bigg[ A_j(y) \dd{}{y^i}
        \rho^{-1}(y) - A_i(x) \dd{}{x^j} \rho^{-1}(x) \bigg].
    \eeq
Hamilton's equation for time evolution of gauge-invariant
observables, $\dot{f} = \{H,f\}$ implies that the magnetic field
evolves according to a non-linear `Euler' equation
    \beq
        \dot{B} = \grad(B/\rho) \times \grad(B \sigma /\rho).
    \eeq
Viewed as a rigid body for the group of volume preserving
diffeomorphisms of $M$, the inverse of the inertia tensor is a
twisted version of the Laplace operator
    \beq
        H = \half \int A_i ~h^{il}~ A_l ~\rho ~d^2x {\rm ~~with~~}
        h^{il} = \fr{\veps^{ij} \veps^{kl}}{\rho} \bigg[ (\pdr_j
        \fr{\sigma}{\rho}) \pdr_k + \fr{\sigma}{\rho} \pdr_j \pdr_k
        \bigg]
    \eeq
This theory has the same phase space and Poisson brackets as $2+1$
ideal(inviscid and volume preserving) hydrodynamics, but a different
hamiltonian. It may be integrable, since we find an infinite number
of conserved quantities $H_n = \int (B/\rho)^n ~\sigma ~\rho ~d^2x $
as well as an infinite number of Casimirs $I_n = \int (B/\rho)^n
~\rho ~d^2x$ for uniform $\rho$. It is remarkable that one can make
this modification to $2+1$ ideal flow, which is sometimes studied as
a toy-model for turbulence, to get a potentially integrable system.
However, our original motivation for studying this model was
different. We argue below that it is the simplest `gauge theory'
that shares some quite deep, though unfamiliar, mathematical
features of Yang-Mills theory.

The formulation of Yang-Mills theory in terms of gauge-invariant
observables, and the development of methods for its solution are
important and challenging problems of theoretical physics, since all
the experimentally observed asymptotic states of the strong
interactions are color-singlets. This problem has a long history
stretching at least as far back as the work of
Mandelstam\cite{mandelstam}. Wilson loops are a natural choice for
gauge-invariant variables, but they have trivial Poisson brackets on
a spatial initial value surface, since the gauge field is
canonically conjugate to the electric field on such a surface. More
recently, it has been shown by Rajeev and
Turgut\cite{rajeev-ictp,rajeev-turgut} , that Wilson Loops of $3+1$
dimensional Yang-Mills theory on a null initial value hypersurface
satisfy a quadratic Poisson algebra with no need for electric field
insertions. This is because the transverse components of the gauge
field satisfy a non-trivial Poisson algebra among themselves, as
opposed to the situation on a spatial surface. The Poisson algebra
of Wilson loops is degenerate due to Mandelstam-like constraints.
The gauge invariant phase space of Yang-Mills theory is conjectured
to be a coadjoint orbit of this Poisson algebra. It is still a
challenge to write the hamiltonian in terms of these variables.
However, this has been possible in dimensionally reduced
versions\footnote{See also the work of Karabali, Nair and Kim who
have made significant progress with a gauge invariant Hamiltonian
approach to $2+1$ Yang-Mills theory\cite{karabali-nair-kim}.} such
as adjoint scalar field theories coupled to quarks in $1+1$
dimensions, as shown by Lee and
Rajeev\cite{lee-rajeev-closed,lee-rajeev-open}. In conjunction with
't Hooft's large $N$ approximation\cite{tHooft-large-N}, viewed as
an alternative classical limit, this is an approach to better
understand the non-perturbative dynamics, especially of
non-supersymmetric gauge theories. In such an approach to $1+1$ QCD,
the phase space of gauge-invariant meson variables is an infinite
Grassmannian, a coadjoint orbit of an infinite dimensional unitary
group\cite{2dqhd}. This allows one to understand baryons as well as
mesons in the large $N$ limit, going beyond the early work of 't
Hooft\cite{tHooft-2dmesons,2dqhd,soliton-parton,gsk-thesis}.

However, the groups and Lie algebras whose coadjoint orbits are
relevant to matrix field theories and Yang-Mills theory are poorly
understood non-commutative versions of diffeomorphism
groups\footnote{This is {\em not} the structure group (sometimes
called the gauge group) of the theory, which is still $SU(N)$ or
$U(N)$. The gauge group plays little role in a gauge-invariant
formulation of the theory.}. In the case of a multi-matrix model,
the group is, roughly speaking, an automorphism group of a tensor
algebra. The Lie algebra is a Cuntz-type algebra which can be
thought of as an algebra of vector fields on a non-commutative
space\cite{rajeev-turgut-der-free-alg,lee-rajeev-closed,lee-rajeev-open,entropy-var-ppl,coll-potn-hamiltonian-mat-mod}.
However, it is still very challenging to find the proper
mathematical framework for these theories and develop approximation
methods to solve them even in the large $N$ limit. To develop the
necessary tools, it becomes worth while to practice on simpler
theories whose gauge invariant phase space is the coadjoint orbit of
a less formidable group. Here, we take a step in this direction by
studying an abelian gauge theory whose phase space is a coadjoint
orbit of the volume preserving diffeomorphism group of a two
dimensional surface.

To put these remarks in perspective, recall the common classical
formulation of Eulerian rigid body dynamics, ideal hydrodynamics,
the KdV equation\cite{arnold-khesin,khesin,kdv-as-euler} and the
large $N$ limit of two dimensional QCD\cite{2dqhd}. The phase space
of each of these theories is a symplectic leaf of a degenerate
Poisson manifold, which is the dual $\gd$ of a Lie algebra. $\gd$
always carries a natural Poisson structure. Symplectic leaves are
coadjoint orbits of a group $G$ acting on the dual of its Lie
algebra $\gd$. On any leaf, the symplectic structure is given by the
Kirillov form. The appropriate groups in these examples are $SO(3)$,
the volume preserving diffeomorphism group of the manifold upon
which the fluid flows, and the central extensions of Diff$(S^1)$ and
of an infinite dimensional unitary group, respectively. The
coadjoint orbits for the rigid body and 2d QCD are well-known
symplectic manifolds: concentric spheres and the infinite
dimensional Grassmannian manifold. The observables in each case are
real-valued functions on $\gd$. The Poisson algebra of observables
is degenerate, i.e. has a center consisting of Casimirs. The
symplectic leaves can also be characterized as the level sets of a
complete set of Casimirs. In each case, the hamiltonian is a
quadratic function on the phase space and classical time evolution
is given by Hamilton's equations. Hamilton's equations are
non-linear despite a quadratic hamiltonian, since the Poisson
brackets of observables are more complicated than the Heisenberg
algebra. In exceptional cases such as the rigid body and the KdV
equation, these non-linear equations are exactly integrable. In
other cases, it is useful to develop approximation methods to solve
them, that are adapted to the geometry of the phase space.

Our earlier remarks indicate that it may be fruitful to regard
Yang-Mills theory and matrix field theories as Hamiltonian dynamical
systems along the lines of the more well known ones listed in the
last paragraph. As a toy-model in this direction, we seek a gauge
theory where the gauge fields satisfy a closed Poisson algebra,
without any need for electric fields. We want a theory whose phase
space is a coadjoint orbit of an ordinary diffeomorphism group,
which is simpler than its non-commutative cousins. We would also
like to understand in more detail the structure of the Poisson
algebra of gauge-invariant observables, work out the equations of
motion and try to solve them.

In this paper, we identify a classical theory of abelian gauge
fields in two spatial dimensions, different from Maxwell theory. In
particular, it is {\em not} Lorentz covariant, indeed, time plays
the same role as in Newtonian relativity. The theory is defined by a
two dimensional manifold $M$, a volume form $\mu$ and a metric
$g_{ij}$ whose volume element $\Omega_g$ is not a multiple of $\mu$.
The hamiltonian is a gauge-invariant magnetic energy, much like that
of Maxwell theory. Unlike in electrodynamics, the gauge field is a
$1$-form on space, rather than on space-time. Thus, {\it even before
any gauge fixing}, the gauge field has no time component. There is a
magnetic field $B$, but no electric field, so to speak. After gauge
fixing, there remains only one dynamical component of the gauge
field. In this sense, the theory has the same number of degrees of
freedom as $2+1$ electrodynamics. However, though the hamiltonian is
quadratic in the gauge fields, the classical theory is nonlinear due
to the `non-canonical' Poisson algebra of gauge fields. Equations of
motion are non-linear and comparable to those of a $2+1$ dimensional
non-abelian gauge theory or ideal hydrodynamics.

The phase space of the theory is a coadjoint orbit of the volume
preserving diffeomorphism group $\sdiff$ of the spatial
two-dimensional manifold. Roughly speaking, this means that $\sdiff$
is a symmetry group of the Poisson algebra of observables. The gauge
group (structure group) of the theory is $U(1)$. The inspiration for
this lies in ideal hydrodynamics. Indeed, even before the
diffeomorphism group of a manifold appeared in general relativity,
it was relevant as the configuration space of a fluid. The theory we
study is not the same as, but is motivated by $2+1$ dimensional
ideal hydrodynamics, regarded as a hamiltonian
system\cite{arnold-inf-dim-groups-hydrodyn,arnold-hamilton-euler-hydro,arnold-khesin,novikov,m-w,d-k-n}.
Though we arrived at it as a toy model for Yang-Mills theory, it
turns out to have a nice geometric and possibly even integrable
structure. We find two infinite sequences of conserved charges. The
first set are Casimirs, analogues of the enstrophy invariants of
ideal hydrodynamics. In addition, we find another infinite set of
conserved charges which are {\em not} Casimirs but are in
involution. The theory we study here can also be regarded as a
theory of geodesics of a right-invariant metric on the volume
preserving diffeomorphism group of a two dimensional manifold.
However, the right invariant metric on $\sdiff$ implied by our
hamiltonian is different from that arising in ideal hydrodynamics
(the $L^2$ metric leading to ideal Euler flow) as well as the $H^1$
metric leading to averaged Euler flow\cite{marsden}.

Another way to view the current work is to recall that adding
supersymmetry usually gives greater analytical control over gauge
theories. But there may be other modifications of gauge theories
that also lead to interesting toy-models or enhanced solvability.
Our investigation concerns one such novel modification of gauge
field Poisson brackets.

In Sec.~\ref{s-vol-pres-vfld-to-gauge-inv-obs} we introduce the
space of abelian gauge fields $A_i dx^i$ on a two dimensional
surface $M$ as the dual of the Lie algebra $\svect$ of vector fields
preserving a volume element $\mu = \rho~ d^2x$. This `duality' is
known in hydrodynamics\cite{arnold-khesin}. The differentials $df^i
= \rho^{-1} \fr{\delta f}{\delta A_i}$ of differentiable
gauge-invariant observables $f(A)$ are shown to be volume preserving
vector fields. In Sec.~\ref{s-pb} we give the Poisson structure on
gauge-invariant observables
    \beq
        \{f,g\} &=& \int d^2x \rho ~ A_i \bigg[ \rho^{-1} \fr{\delta f}{\delta A_j}
            \pdr_j \bigg( \rho^{-1} \fr{\delta g}{\delta A_i} \bigg)
            - \rho^{-1} \fr{\delta g}{\delta A_j} \pdr_j \bigg(
           \rho^{-1} \fr{\delta f}{\delta A_i} \bigg) \bigg]
    \eeq
turning the space of gauge fields into a Poisson manifold. The
Poisson brackets of gauge fields are obtained explicitly
(\ref{e-pb-of-gauge-flds}) and compared with those of Yang-Mills
theory on a spatial and null initial value hypersurface.

In Sec.~\ref{s-str-of-poisson-alg} we give the coadjoint action of
$\sdiff$ and its Lie algebra $\svect$ on the Poisson manifold of
gauge fields $\svect^*$, and show that the action is canonical i.e.
preserves the Poisson structure. The moment maps generate the
coadjoint action. The symplectic leaves of the Poisson manifold are
coadjoint orbits. The enstrophy invariants of hydrodynamics $I_n =
\int_M (dA/\mu)^n \mu$ are an infinite sequence of Casimirs of the
Poisson algebra. The coadjoint orbits of closed gauge field
$1$-forms are shown to be finite dimensional. Single-point orbits
for simply connected $M$ are found. We argue that all other orbits
are infinite dimensional and try to characterize their isotropy
subalgebras as well as tangent spaces.

In Sec.~\ref{s-hamiltonian} we first review the choice of
hamiltonian leading to ideal Eulerian hydrodynamics in 2+1
dimensions. Then we propose a different gauge-invariant hamiltonian
depending on both $\mu$ and a metric $g_{ij}$, by analogy with the
magnetic energy of Maxwell theory.
    \beq
        H = \half \int_M \bigg( \fr{F \wedge *F}{\Omega_g}\bigg)
        \mu = \int \bigg(\fr{B}{\rho} \bigg)^2 ~\sig ~\rho~ d^2x
    \eeq
where $F=dA$ is the field strength, $*F$ is its Hodge dual, and
$\Om_g$ is the volume element of the metric $g_{ij}$. Here $\sigma =
(\mu/\Om_g)^2 = \rho^2/g$, $g = \det{g_{ij}}$ and $\rho$ is the
density associated to $\mu$. $H$ is shown to determine a
non-negative inner product on the dual of the Lie algebra $\svect^*$
and an inverse `inertia tensor' by analogy with the rigid body. If
$M$ is simply connected, the inverse inertia operator is
non-degenerate and could be inverted to get an inner product on the
Lie algebra $\svect$. This could be extended to the diffeomorphism
group $\sdiff$ by right translations. Thus, the magnetic energy
should define geodesic flow on $\sdiff$ with respect to a
right-invariant metric different from that coming from Eulerian
hydrodynamics.

In Sec.~\ref{s-eqn-motion} we find the equation of motion for the
magnetic field $B = \veps^{ij} \pdr_i A_j$, $\dot{B} = \grad(B/\rho)
\times \grad(B \rho /g)$. This simple quadratically non-linear
evolution equation is strikingly similar to the Euler equation of a
rigid body $\dot{L} = L \times \Om,~ L = I \Om$. It can be regarded
as the `Euler equation' for the group $\sdiff$ with hamiltonian
given above. Remarkably, for a uniform measure $\mu$ we find an
infinite sequence $H_n$ of conserved charges in involution, which
are not Casimirs. The hamiltonian is $\fr{H_2}{2}$.
    \beq
        H_n = \int_M (dA/\mu)^n \sigma \mu, ~~~ n = 1,2,3, \ldots
    \eeq
In Sec.~\ref{s-static-solns} we find some static solutions of the
equations of motion. We show that for circularly symmetric $\rho$
and $g$, every circularly symmetric magnetic field is a static
solution. We generalize this to the non-symmetric case as well. We
also find a one parameter family of static solutions that are local
extrema of energy even with respect to variations that are not
restricted to the symplectic leaf on which the extremum lies. Some
ideas for further study are given in Sec.~\ref{s-discussion}.

Volume preserving diffeomorphisms and gauge theories have appeared
together previously in the literature (see for example
Ref.~\cite{unrelated-but-similar-sounding-1,unrelated-but-similar-sounding-2}).
Our investigation seems quite different, since $\sdiff$ is {\em not}
the gauge group of our theory but rather a symmetry of the Poisson
algebra.

\section{Volume Preserving Vector Fields to Gauge Invariant Observables}
\label{s-vol-pres-vfld-to-gauge-inv-obs}

\subsection{Lie Algebra of Volume Preserving Vector Fields}

Let $M$ be a surface with local coordinates $x^i$, to be thought of
as the space on which a fluid flows. A vector field on $M$ is
regarded as the velocity field of a fluid at a particular time. The
space of all vector fields on $M$ forms a Lie algebra $\vect$ with
Lie bracket
    \beq
        [u,v]^i = u^j \pdr_j v^i - v^j \pdr_j u^i
    \eeq
$\vect$ is the Lie algebra of the diffeomorphism group $\diff$.
Conservation of the mass of the fluid during its flow implies the
continuity equation for its density $\rho(x,t)$
    \beq
        \dd{\rho(x,t)}{t} + \grad \cdot (\rho {\bf u}) = 0
    \eeq
We are interested in flows where the density at any point of space
does not depend on time. Using the continuity equation, this becomes
$\nabla \cdot (\rho {\bf u}) = 0$. We call such a flow volume
preserving. Geometrically, we are considering a flow that generates
diffeomorphisms of $M$ that preserve a given volume form\footnote{A
volume form must be non-degenerate. In $2$ dimensions it is the same
as an area form or a symplectic form.} ${\cal L}_u \mu = 0$. The
density is constant along integral curves of $u$. To see the
equivalence of this with the continuity equation for a volume
preserving flow, recall that
    \beq
        {\cal L}_u \mu = (d i_u + i_u d) \mu = d (i_u \mu)
    \eeq
where $i_u$ is the contraction with $u$. Here $d\mu = 0$ since $\mu$
is a volume form. In local coordinates $\mu = \half \mu_{ij} dx^i
\wedge dx^j$ where $\mu_{ij} = - \mu_{ji} \equiv \eps_{ij} \rho$ and
$\eps_{ij}$ is antisymmetric with $\eps_{12} = 1$. So $\mu = \rho(x)
~ d^2x$ where $dx^1 \wedge dx^2 \equiv d^2x$. Then $i_u \mu = \half
\mu_{ij} (u^i dx^j - dx^i u^j) = \mu_{ij} u^i dx^j$, so that
    \beq
    {\cal L}_u \mu &=& d (i_u \mu)  = \pdr_k (\mu_{ij} u^i)
        dx^k \wedge dx^j = \pdr_i (\rho u^i) dx^1 \wedge dx^2
    \eeq
Thus ${\cal L}_u \mu = 0$ becomes $ \grad \cdot (\rho {\bf u}) = 0$.
We will use the terms volume preserving and area preserving
interchangeably since $M$ is a two dimensional surface. Some of what
we say has a generalization to higher (especially even) dimensional
$M$.

The properties ${\cal L}_{\alpha u + \beta v} = \alpha {\cal L}_{u}
+ \beta {\cal L}_{v}$ and ${\cal L}_{[u,v]} = {\cal L}_u {\cal L}_v
- {\cal L}_v {\cal L}_u$ ensure that the space of volume-preserving
vector fields $\g = \svect$ forms a Lie subalgebra of $\vect$. It is
the Lie algebra of the group of volume preserving diffeomorphisms $G
= \sdiff$.

Volume preserving flow is a mathematical implementation of the
physical concept of incompressible flow, which occurs when the fluid
speed is small compared to the speed of sound\footnote{Some authors
consider only the special case where density is a constant, $\nabla
\cdot u=0$. Note also that the same fluid may support both
compressible and incompressible flow under different conditions, so
our definitions refer to the flow and not just to the fluid.}. In
particular, shock waves cannot form in incompressible flow, since
shock waves involve supersonic flow. Under ordinary conditions, air
flow in the atmosphere is incompressible. Volume preserving flow is
sometimes referred to as divergence-free flow.

If $M$ is simply connected, the volume preserving condition $\pdr_i
(\rho u^i) = 0$ may be solved in terms of a {\em stream function}
$\psi$ satisfying $i_u \mu = d\psi$. $\psi$ is a scalar function on
$M$ that serves as a `potential' for the velocity field. In local
coordinates
    \beq
        i_u \mu = d\psi \Rightarrow \mu_{ij} u^i = \pdr_j \psi
    \eeq
Since $\mu$ is non-degenerate ($\rho \ne 0$), it can be inverted
$\rho^{-1} \veps^{ij} \mu_{jk} = - \delta^i_k$ where $\veps^{ij}$ is
a constant antisymmetric tensor with $\veps^{12} = 1$, $\veps^{ij}
\eps_{jk} = - \delta^i_k$. This does not require a metric on $M$.
Then
    \beq
        u^i = \rho^{-1} \veps^{ij} \pdr_j \psi.
    \eeq
$u$ determines $\psi$ up to an additive constant, which can be fixed
by a boundary condition. If $M$ is simply connected, then $\svect$
may be identified with the space of stream functions. Suppose two
volume preserving vector fields $u,v$ have stream functions $\psi_u$
and $\psi_v$,
    \beq
    u^i = \rho^{-1} \veps^{ij} \pdr_j \psi_u, ~~ v^i = \rho^{-1} \veps^{ij} \pdr_j \psi_u
    \eeq
Then their Lie bracket $[u,v]$ has stream function $\rho^{-1} \grad
\psi_v \times \grad \psi_u$:
    \beq
        [u,v]^i &=& \rho^{-1} \veps^{il} \pdr_l \bigg\{ \rho^{-1}
            \veps^{jk} (\pdr_j \psi_v) (\pdr_k \psi_u) \bigg\} \cr
        \psi_{[u,v]} &=& \rho^{-1} \veps^{jk} (\pdr_j \psi_v) (\pdr_k \psi_u)
    \eeq

\subsection{Abelian Gauge Fields as the Dual of $\svect$}

The Lie algebra $\svect$ is akin to the Lie algebra of angular
velocities of a rigid body. The angular momenta are in the dual
space to angular velocities, and satisfy the angular momentum
Poisson algebra. As explained in
\ref{a-dual-of-lie-alg-coadjoint-orbits}, the dual of any Lie
algebra is a Poisson manifold. This is interesting because the
observables of a classical dynamical system are real-valued
functions on a Poisson manifold. The dual of the Lie algebra $\g =
\svect$ is the space of abelian gauge fields modulo gauge
transformations,
    \beq
    \gd = \svect^* = \Omega^1(M) / d \Omega^0(M).
    \eeq
This fact is well-known in hydrodynamics (see
Ref.~\cite{arnold-khesin}), though it is usually not thought of in
terms of gauge fields. To see this duality, we define the pairing
$(A,u)$ between gauge fields and volume preserving vector fields by
integrating the scalar $A(u)$ with respect to $\mu$
    \beq
    (A,u) = \mu_u(A) = \int_M A(u) \mu = \int A_i u^i ~ \rho ~ d^2x
    \eeq
The pairing $\mu_u(A)$ is also called the moment map. It is a
gauge-invariant pairing. Under a gauge transformation $A \mapsto A'
= A + d \La$ for any scalar $\La(x)$
    \beq
    \mu_u(A') - \mu_u(A) = \int_M (\pdr_i \La) u^i \mu = - \int \La
    \pdr_i(\rho u^i) d^2x = 0
    \eeq
since $u$ is volume preserving. We assume that gauge fields and
gauge transformations $\La$ vanish on the boundary $\pdr M$ or at
infinity. We make no such assumption about the vector fields.

Gauge Fixing: It is occasionally convenient to `gauge-fix', i.e.
pick a coset representative for $\Om^1(M)/d\Om^0(M)$. Under a gauge
transformation, $A_i^{\prime} = A_i + \pdr_i \La$. We can pick $\La$
such that $A_1^{\prime} = A_1 + \pdr_i \La = 0$, so that we are left
with only one component of the gauge field $A_2^\prime$. We can
still make an $x^1$-independent `residual' gauge transformation,
$A_2^{\prime \prime} = A_2^\prime + \pdr_2 \tilde \La(x^2)$ to
eliminate any additive term in $A_2^\prime$ depending on $x^2$
alone. Suppose we have gauge fixed on a particular spatial initial
value surface at time $t=0$. Unlike in Yang-Mills theory, the
equations of motion of our theory are purely dynamical. They only
evolve the gauge-fixed fields forward in time, and do not contain
any further constraints. In effect, after gauge fixing, we will be
left with one propagating field degree of freedom.

\subsection{Differentials of Gauge Invariant Charges are
Volume Preserving Vector Fields}
\label{s-diff-of-gauge-inv-are-vol-pres}

We should regard $\svect^*$, the space of gauge fields modulo gauge
transformations, as the Poisson manifold of some dynamical system.
Real-valued functions on this space (i.e. gauge-invariant functions
$f(A)$) are the observables. Given such an $f(A)$, we can define its
differential $df(A)$
    \beq
    (df(A))^i = \rho^{-1}(x) \fr{\delta f}{\delta A_i(x)} \equiv
        \rho^{-1} \delta^i f
    \label{e-defn-of-differential}
    \eeq
For each equivalence class of gauge fields $[A] = \{A | A \sim A + d
\La \}$, the differential\footnote{The differential $(df(A))^i$ is
regarded as a vector field on $M$ for each $A$ and should not be
confused with the closely related exterior derivative $df$, which is
a 1-form on $\gd$. However, as we will see later
(Sec.~\ref{s-str-of-poisson-alg},
\ref{a-dual-of-lie-alg-coadjoint-orbits}), on any symplectic leaf of
$\svect^*$ with symplectic form $\om$, the $1$-form $df$ determines
the canonical vector field $V_f$ via $\om(V_f,.) = df(.)$. $V_f$ is
a vector field on the leaf, and its relation to the differential is
$V_f(A) = ad^*_{df} A$.} defines a vector field $df^i \pdr_i$ on
$M$. If $f(A)$ is non-linear, the vector field $(df(A))^i(x)$
changes as $A \in \gd$ changes. Suppose $f(A)$ is gauge invariant
and differentiable. Then we can show that its differential $df^i$ is
a volume preserving vector field on $M$: $\pdr_i(\rho ~df^i) = 0$.
To see this, note that gauge invariance implies that the change in
$f$ under any gauge transformation $\delta A_i = \pdr_i \La$ must
vanish
    \beq
    0 = \delta f = \int \fr{\delta f}{\delta A_i(x)} \delta A_i(x)
        d^2x = \int \fr{\delta f}{\delta A_i(x)} \pdr_i \La
        d^2x = - \int \pdr_i \bigg(\fr{\delta f}{ \delta A_i(x)}
        \bigg) \La(x) d^2x
    \eeq
Since $\La(x)$ is arbitrary, it must follow that $\pdr_i(\delta f /
\delta A_i) =0$. So the differential of a gauge invariant function
can be regarded as an element of the Lie algebra $\g = \svect$.

The simplest gauge-invariant observable is the field strength
$2$-form
    \beq
    F = dA = \half F_{ij} dx^i \wedge dx^j; ~~~ F_{ij} = \pdr_i A_j
        - \pdr_j A_i = \eps_{ij} B
    \label{e-defn-of-fld-strength-mag-fld}
    \eeq
where $B = \veps^{ij} \pdr_i A_j = \pdr_1 A_2 - \pdr_2 A_1$ is the
magnetic field. The differential of $F$
    \beq
    ((dF)(A))^k =  \rho^{-1} \fr{\delta F_{ij}(x)}{\delta A_k(y)} =
        \rho^{-1} \bigg( \delta^k_j \pdr_i \delta(x-y)
        - \delta^k_i \pdr_j \delta(x-y) \bigg)
    \eeq
is a volume preserving vector field on $M$ for each $A$:
    \beq
    \pdr_k (\rho (dF)^k) =  \pdr_k \bigg( \delta^k_j \pdr_i \delta(x-y)
        - \delta^k_i \pdr_j \delta(x-y) \bigg) = (\pdr_i \pdr_j - \pdr_j
        \pdr_i) \delta(x-y) = 0.
    \eeq
Similarly, the differential of the magnetic field
    \beq
    ((dB)(A))^k = \rho^{-1} \fr{\delta B(x)}{\delta A_k(y)} = \rho^{-1}
         \veps^{ij} \pdr_i \delta^k_j \delta(x-y)
    = \rho^{-1}  \veps^{ik} \pdr_i \delta(x-y)
    \eeq
is volume preserving $\pdr_k(\rho (dB)^k) =  \veps^{ik} \pdr_i
\pdr_k \delta(x-y) = 0$. We can regard the moment maps $\mu_u(A)$ as
linear gauge-invariant observables. The differential of $\mu_u(A)$
is the volume preserving vector field $u$, for all gauge fields $A$.

Other gauge invariant observables $f(A)$ we will be interested in
are `charges': integrals over $M$ with respect to $\mu$, of a local
gauge-invariant scalar function ${\cal F}$. $\cal F$ can depend on
$A$ only through the field strength two form $F = dA$. The analogue
of the Chern-Simons 3-form, vanishes identically since $A$ is a
$1$-form on space, not space-time. The quotient of $dA$ and the
non-degenerate volume 2-form $(dA/\mu)$ is a scalar function on $M$.
Then
    \beq
    f(A) = \int {\cal F}(\sigma,~ (dA/\mu),~ v^i \pdr_i(dA/\mu),
        ~ w^{ij} \pdr_i \pdr_j(dA/\mu),~ \ldots)~~ \mu
    \eeq
where $\sig$ is a scalar function and $v^{i}, w^{ij}$ etc are
arbitrary but fixed contravariant tensor fields. We can get an
explicit formula for the differential of such a gauge invariant
charge. Using $dA/\mu = B/\rho$ and
    \beq
    \dd{B}{A_i} = - \veps^{ij} \pdr_j \delta^2(x-y); &&
    \fr{\delta \pdr_k(B/\rho)(x)}{\delta A_i(y)} = - \veps^{ij}
    \pdr_k\bigg( \ov{\rho} \pdr_j \delta^2(x-y) \bigg); \ldots
    \eeq
we get upon integrating by parts,
    \beq
        df(A)^i = \rho^{-1} \fr{\delta f}{\delta A_i} = \rho^{-1} \veps^{ij}
            \pdr_j \bigg[\bigg(\dd{{\cal F}}{(B/\rho)}
            \bigg) - \bigg( \ov{\rho} \pdr_k \bigg(
            \dd{\cal F}{\pdr_k(B/\rho)}
            \bigg)  \bigg) + \ldots  \bigg]
    \eeq
Due to the anti-symmetry of $\veps^{ij}$, it follows that $df$ is
volume preserving $\pdr_i(\rho~ df^i) = 0$. Moreover, if $f$ is
gauge-invariant and of the form assumed above, then its differential
is also gauge-invariant.

Two families of gauge invariant charges which play an important role
in our theory are $I_n$ and $H_n$ defined below. Let
    \beq
        I_n(A) &=& \int_M (dA/\mu)^n ~\mu  = \int (B / \rho)^n ~ \rho
        ~d^2x = \int_M \bigg(\fr{B}{\rho}\bigg)^{n-1}~ dA, ~~~
        n=1,2,3,\ldots  \cr
        (dI_n)^i &=& \ov{\rho} \fr{\delta I_n}{\delta A_i}
        =  \ov{\rho} n  \veps^{ij} \pdr_j ((B/\rho)^{n-1})
    \label{e-differential-of-In}
    \eeq
Their differentials are volume preserving $\pdr_i (\rho (dI_n(A))^i)
= 0$ since $\veps^{ij}$ is antisymmetric. Note that $I_1 = \int_M dA
= 0$. We assume $B$ vanishes sufficiently fast at infinity and do
not consider $I_n$ for $n < 0$. Given a scalar function $\sigma$ on
$M$ we can construct additional gauge-invariant charges. These are
similar to the $I_n$, except that we multiply by $\sigma$ before
integrating over $M$
    \beq
        H_n(A) = \int (dA/\mu)^n \sig \mu = \int (B/\rho)^n
        ~\sig ~\rho ~d^2x
    \eeq
More generally, one can replace $(B/\rho)^n$ by an arbitrary
function of $B/\rho$. The differential of $H_n$ is volume
preserving:
    \beq
        (dH_n(A))^i = \fr{n  \veps^{ij}}{\rho} \pdr_j(\sig
        (B/\rho)^{n-1}).
    \eeq
Gauge-invariant observables depending on the volume form $\mu$ and a
metric $g_{ij}$ on $M$ will play an important role in determining
the dynamics of our theory. They are given in
Sec.~\ref{s-hamiltonian-as-mag-egy}.

There are other interesting gauge invariant observables such as the
circulation $C_\gamma(A) = \int_0^1 A_i \fr{d\gamma^i(s)}{ds} ds$
and its exponential, the Wilson loop. Since these observables are
concentrated on one dimensional curves on $M$, they may fail to be
differentiable and their differentials exist only as distributional
vector fields. They require a more careful analysis and are not
considered in this paper. Henceforth, when we say observable, we
will mean gauge invariant observables that are differentiable, in
which case, their differentials are guaranteed to be volume
preserving vector fields on $M$.

\section{Poisson Brackets}
\label{s-pb}

\subsection{Definition of Poisson Bracket on $\gd = \svect^*$}

The Lie algebra structure of volume preserving vector fields $\g =
\svect$ can be used to define a Poisson structure on its dual (see
\ref{a-dual-of-lie-alg-coadjoint-orbits},
Ref.~\cite{kirillov,weinstein-poisson,arnold-class-mech}). The dual
space $\gd = \svect^* = \Omega^1(M)/d\Omega^0(M)$ of gauge field
1-forms then becomes a Poisson manifold. Observables are real-valued
functions on it. The Poisson bracket (p.b.) between gauge-invariant
observables $f(A)$ and $g(A)$ with volume preserving differentials
$df$ and $dg$, is defined using the pairing $(A,u)$ between $\g$ and
$\gd$ and the Lie algebra bracket $[df,dg]$:
    \beq
        \{f,g\}(A) &\equiv& (A,[df,dg]) = \int_M A_i [df,dg]^i
        \mu = \int d^2x \rho A_i [df^j \pdr_j dg^i - dg^j \pdr_j
            df^i] \cr
        &=& \int d^2x ~\rho ~A_i \bigg[ \rho^{-1} \fr{\delta f}{\delta A_j}
            \pdr_j \bigg( \rho^{-1} \fr{\delta g}{\delta A_i} \bigg)
            - \rho^{-1} \fr{\delta g}{\delta A_j} \pdr_j \bigg(
            \rho^{-1} \fr{\delta f}{\delta A_i} \bigg) \bigg] \cr
        &=& \int A_i \bigg[(\delta^j f)
            \pdr_j ( \rho^{-1} \delta^i g )
            -  (\delta^j g) \pdr_j ( \rho^{-1}
            \delta^i f )\bigg] ~d^2x
    \label{e-pb-formula}
    \eeq
where $\delta^i = \fr{\delta}{\delta A_i}$. The antisymmetry,
linearity and Jacobi identity follow from the corresponding
properties of the Lie bracket. The Leibnitz rule follows from the
Leibnitz rule for differentials.

The p.b.(\ref{e-pb-formula}) preserves the class of gauge-invariant
functions. Suppose $f,g$ are gauge-invariant. Then $df, dg \in
\svect$. Recall (Sec.~\ref{s-diff-of-gauge-inv-are-vol-pres}) that
if $f$ and $g$ are gauge-invariant, then so are their differentials
$df^i$ and $dg^i$. It follows that $[df,dg]^i$ is also
gauge-invariant. Now, under a gauge transformation $A' = A + d\La$,
    \beq
    \{f,g\}(A') &=& (A',[df,dg](A')) = (A',[df,dg](A)) = \int A'_i
        ([df,dg](A))^i \mu \cr
    \Rightarrow && \{f,g\}(A') - \{f,g\}(A) = - \int \La \pdr_i(
        ([df,dg](A))^i \rho) d^2x = 0
    \eeq
Thus (\ref{e-pb-formula}) is a gauge invariant Poisson bracket.

A gauge-invariant observable $f(A)$ defines canonical
transformations on the Poisson manifold $\gd$ via the p.b. Suppose
$g(A)$ is any observable, then its Lie derivative under the flow
generated by $f$ is ${\cal L}_{V_f} g(A) = \{f,g\}(A) = (A,[df,dg])$
(see Sec.~\ref{s-str-of-poisson-alg}).

\noindent {\bf Example:} The p.b. of two moment maps $\mu_u(A)$ and
$\mu_v(A)$ is
    \beq
        \{\mu_u , \mu_v\}(A) = \mu_{[u,v]}(A) = \int_M A_i (u^j \pdr_j v^i - v^j \pdr_j u^i)
        \mu
    \eeq
If $u$ and $v$ are volume preserving, then so is $[u,v]$; therefore,
if $\mu_u$ and $\mu_v$ are gauge-invariant functions of $A$, then so
is $\mu_{[u,v]}$. Moreover, the canonical transformation generated
by the moment map $\mu_u$ is just the Lie algebra coadjoint action
(see \ref{a-dual-of-lie-alg-coadjoint-orbits} and
Sec.~\ref{s-str-of-poisson-alg})
    \beq
    {\cal L}_{V_{\mu_u}} f(A) = \{\mu_u, f\}(A); ~~~~
        {\cal L}_{V_{\mu_u}} A = ad^*_u  A.
    \eeq

\subsection{Poisson Brackets of Gauge Fields}

Equivalence classes of gauge fields are coordinates on our Poisson
manifold $\gd = \svect^*$. Thus, an explicit formula for the
`fundamental' p.b. between components of the gauge field $A_i(x)$ is
useful. This will also facilitate a comparison with electrodynamics.
We will show that the p.b. between gauge field components is
    \beq
        \{A_i(x),A_j(y) \} = \delta^2(x-y) \bigg[ A_j(y) \dd{}{y^i}
        \rho^{-1}(y) - A_i(x) \dd{}{x^j} \rho^{-1}(x) \bigg],
        \label{e-pb-of-gauge-flds}
    \eeq
where the derivatives act on everything to their right. Though this
formula for $\{A_i(x),A_j(y)\}$ looks a bit complicated, the rhs is
linear in gauge fields. Thus, our Poisson algebra is actually a Lie
algebra like the Lie algebra of angular momenta  $\{L_i, L_j \} =
\eps_{ijk} L_k$.

Recall that electrodynamics is based on the Heisenberg algebra
between the spatial components of the gauge field and the spatial
components of the electric field. In two spatial dimensions we have
(before any gauge fixing)
    \beq
        \{A_i({\bf x},t),E^j({\bf y},t) \} &=& \delta^2({\bf x}-{\bf
        y}) \delta_i^j \cr
        \{A_i({\bf x},t),A_j({\bf y},t) \} &=& 0 \cr
        \{E^i({\bf x},t),E^j({\bf y},t) \} &=& 0
    \eeq
While in electrodynamics the electric field is canonically conjugate
to the gauge field, this is {\em not} the case in our theory. The
components of the gauge field in our theory obey p.b. relations with
each other, without being canonically conjugate. This is not
unusual. For instance, the components of angular momentum form a
closed Poisson algebra though none of them is canonically conjugate
to another. This is a generic feature of degenerate Poisson
manifolds where canonically conjugate variables can only be chosen
on individual symplectic leaves (the concentric spheres in the case
of angular momenta).

Gauge fields obeying p.b. not involving the electric field are {\em
not} alien to conventional Yang-Mills theory. For example, in a
coordinate system where initial values of fields are specified on a
null cone at past time-like infinity, the transverse components of
gauge fields satisfy p.b. among themselves as shown by Rajeev and
Turgut\cite{rajeev-ictp,rajeev-turgut}
    \beq
        \{A_{ib}^a(z,R), A_{jd}^c(z',R') \} = \half \delta^a_d
        \delta^c_b q_{ij}(z) \delta(z-z') \sgn(R-R')
    \eeq
where $z^i$ are transverse angular coordinates, $q_{ij}$ is the
round metric on $S^2$, $R$ is a radial coordinate and $a,b$ are
color indices. In fact, our original motivation for studying the
dynamical system in this paper was to find a toy-model that shared
this feature with Yang-Mills theory.

Now we will establish (\ref{e-pb-of-gauge-flds}) using
(\ref{e-pb-formula}) and the relation between the p.b. of functions
and those between the `coordinates' $A_i(x)$
    \beq
        \{f,g\}(A) = \int d^2x d^2y \{A_i(x), A_j(y)\}
        \fr{\delta f}{\delta A_i(x)} \fr{\delta g}{\delta A_j(y)}
    \eeq
We rewrite $\{f,g \}(A)$ from (\ref{e-pb-formula}) to make it look
like this
    \beq
    && \int d^2x A_j \bigg[ \fr{\delta f}{\delta A_i}
            \pdr_i \bigg( \ov{\rho} \fr{\delta g}{\delta A_j} \bigg)
            - \fr{\delta g}{\delta A_i} \pdr_i \bigg( \ov{\rho}
            \fr{\delta f}{\delta A_j} \bigg) \bigg] \cr
        &=& \int d^2x d^2y \delta^2(x-y) A_j(y) \bigg[ \fr{\delta f}{\delta
            A_i(x)} \dd{}{y^i} \bigg(\ov{\rho(y)} \fr{\delta g}{\delta A_j(y)}
            \bigg) - f \leftrightarrow g \bigg] \cr
        &=& \int d^2x d^2y \delta^2(x-y) A_j(y) \dd{}{y^i}
            \bigg[\ov{\rho(y)} \fr{\delta f}{\delta A_i(x)} \fr{\delta g}{\delta
            A_j(y)} - f \leftrightarrow g \bigg] \cr
        &=& \int d^2x d^2y \delta^2(x-y) \bigg[ A_j(y) \dd{}{y^i}
            \rho^{-1}(y) - A_i(x) \dd{}{x^j} \rho^{-1}(x) \bigg]
            \fr{\delta f}{\delta A_i(x)} \fr{\delta g}{\delta
            A_j(y)}.
    \eeq
Finally we read off the desired expression
(\ref{e-pb-of-gauge-flds}).

\section{Structure of Poisson Algebra of Observables}
\label{s-str-of-poisson-alg}

In this section we give the canonical action of $\sdiff$ (generated
via p.b.) on the Poisson manifold $\gd = \svect^*$ (see also
\ref{a-dual-of-lie-alg-coadjoint-orbits},
Ref.~\cite{arnold-khesin}). Symplectic leaves are coadjoint orbits
of $\sdiff$. The Poisson algebra of gauge invariant observables is
degenerate. The enstrophy invariants $I_n$ of hydrodynamics are an
infinite number of Casimirs. They are constant on the coadjoint
orbits. Then we try to identify some of the simpler coadjoint
orbits. We show that there is always at least one single-point
orbit, that of the pure gauge configuration. If $M$ has
non-vanishing first cohomology, then we show that there are finite
dimensional symplectic leaves lying inside $H^1(M) \setminus \{0\}$.
This also shows that $I_n$ could not be a complete set of coadjoint
orbit invariants. If $M$ is simply connected, we show that the only
single-point orbits consist of the configurations for which $dA/\mu$
is constant. For simply connected $M$, we also argue that the orbit
of $[A] \in \gd$ for which $dA/\mu$ is not constant, is infinite
dimensional. We identify the isotropy sub-algebra and tangent space
of such an orbit and give an example where $dA/\mu$ is circularly
symmetric. However, this analysis is far from complete. It would be
useful to find a nice coordinate system on these orbits and get the
symplectic structure in explicit form so as to study hamiltonian
reduction.

\subsection{Coadjoint Action of $\sdiff$ on Poisson Algebra}
\label{s-coadj-action-of-sdiff-on-poiss-alg}

The Poisson manifold $\gd = \Omega^1(M) / d \Omega^0(M) = \{A \in
\Om^1(M) ~| A \sim A + d\La\}$ carries the coadjoint action of
$\sdiff$. For example, on a simply connected region $M$, $\gd$ may
be identified with the space of scalar functions $f = (dA / \mu) =
(B / \rho)$ on $M$. The coadjoint action is the pull-back action of
volume preserving diffeomorphisms $\phi \in \sdiff$ on functions
$\phi^* f = f \circ \phi$. The action on equivalence classes of
gauge fields is also the pull back $Ad_\phi^*[A] = [\phi^* A]$. For
infinitesimal $\phi(t) = 1 + ut$ we get the Lie algebra coadjoint
action of $u \in \svect$ on $\gd$
    \beq
        ad^*_u A = - {\cal L}_u A
    \eeq
This action of $\sdiff$ on $\gd$ is canonical i.e. there is a
function on $\gd$ (a gauge-invariant observable) that generates the
coadjoint action via the p.b. The generating function is the moment
map $\mu_u(A)$. To see this, suppose $u$ is a volume preserving
vector field, then
    \beq
        \{\mu_u(A), A_j \} = -({\cal L}_u A)_j.
    \label{e-mom-maps-gen-coadj-action}
    \eeq
To show this we begin with the rhs of
(\ref{e-mom-maps-gen-coadj-action})
    \beq
    {\cal L}_u A &=& d(i_u A) + i_u(dA)
        = d(u^i A_i) + \half i_u \bigg\{(\pdr_j A_i - \pdr_i A_j) dx^i \wedge
        dx^j \bigg\}  \cr
    &=& (\pdr_j A_i u^i + A_i \pdr_j u^i) dx^j + \half (\pdr_j A_i - \pdr_i
        A_j) (u^j dx^i - dx^j u^i) \cr &=& \bigg\{ A_i \pdr_j u^i + u^i \pdr_i A_j \bigg\} dx^j
    \eeq
On the other hand, using (\ref{e-pb-of-gauge-flds}),
$\{\mu_u(A),A_k(z)\}$ is equal to
    \beq
         && \int d^2x ~d^2y ~\{A_i(x),A_j(y)\} ~\fr{\delta \mu_u}{
            \delta A_i(x)} ~\fr{\delta A_k(z)}{\delta A_j(y)} \cr
        &=& \int d^2x ~d^2y ~ \delta^2(x-y) \bigg[A_j(y) \dd{}{y^i} \rho^{-1}(y)
            - A_i(x) \dd{}{x^j} \rho^{-1}(x) \bigg] \rho(x) u^i(x)
            \delta^j_k \delta^2(z-y) \cr
        &=& \int d^2y~ A_k(y) \rho(y) u^i(y) \dd{}{y^i} (\rho^{-1}(y) \delta^2(y-z))
            - \int d^2x~ A_i(x) \delta^2(z-x) \dd{u^i(x)}{x^k} \cr
        &=& - \rho^{-1} \pdr_i (A_k \rho u^i) - A_i \pdr_k u^i
        = - u^i \pdr_i A_k - A_i \pdr_k u^i = - ({\cal L}_u A)_k
    \eeq
We integrated by parts ($A = 0$ on $\pdr M$) and used $\pdr_i(\rho
u^i)=0$. Thus, we obtain (\ref{e-mom-maps-gen-coadj-action}).

\subsection{Center of Poisson Algebra}

From Sec.~\ref{s-coadj-action-of-sdiff-on-poiss-alg}, the symplectic
leaves of $\svect^*$ are homogeneous symplectic manifolds,
identified with coadjoint orbits of $\sdiff$. Is the Poisson algebra
of functions on $\svect^*$ degenerate? What are the Casimirs that
constitute its center? Casimirs are unchanged under canonical
transformations. So they are constant on symplectic leaves. A
complete set of such Casimirs would allow us to distinguish between
distinct leaves. Since the symplectic leaves are coadjoint orbits of
$\sdiff$, it suffices to find observables that commute with the
moment maps $\mu_u(A)$, which generate the coadjoint action.

We first observe that the center of the Poisson subalgebra of linear
observables $\mu_u(A)$ is trivial. $\{\mu_u, \mu_v\}(A)$ vanishes
for all $A$ iff $[u,v]=0$. But the center of $\svect$ is trivial.
This is seen by writing $u$ and $v$ in terms of their stream
functions $\psi_u, \psi_v$. Taking some simple choices for $\psi_v$
in the condition $[u,v]=0$ will imply that $\psi_u = 0$. Thus none
of the $\mu_u(A)$ lie in the center of the Poisson algebra of
gauge-invariant functions, since they do not even commute with each
other. Thus we need to look elsewhere to find the Casimirs of our
Poisson algebra. It turns out that the charges $I_n = \int
(dA/\mu)^n \mu$ are central observables. The monomials $(dA/\mu)^n$
can be replaced by any scalar function of $(dA/\mu)$. We found that
$I_n$ are Casimirs by showing that they Poisson commute with the
moment maps. However, this involves lengthy calculations. For eg.,
in \ref{a-In-are-in-involution} we show that $I_n$ commute with each
other and in \ref{a-I2-lies-in-center} we show that $I_2$ commutes
with $\mu_u(A)$ for uniform $\mu$.

However, we found that $I_n$ are closely related to the enstrophy
invariants of hydrodynamics, which are known to be conserved
quantities in Eulerian hydrodynamics. There is a simple argument in
Ref.~\cite{arnold-khesin} based on earlier
work\cite{m-w,novikov,d-k-n} that leads to the conclusion that $I_n$
are constant on coadjoint orbits. The argument is that the action of
$\sdiff$ is merely to change coordinates in the integral defining
$I_n$. Since this integral is independent of the choice of
coordinates, $I_n$ must be invariant under the coadjoint action.
Though $I_n$ for $n = 1,2,3,\ldots$ are all Casimirs, they are not a
complete set of orbit invariants. Their level sets do not
necessarily distinguish the coadjoint orbits (see
Sec.~\ref{s-orbits-of-closed-forms}). For some remarks on additional
invariants, see  Sec.~9 of Ref.~\cite{arnold-khesin}. Our explicit
calculations of $\{I_n,\mu_u\}$ given in
\ref{a-In-are-in-involution} and \ref{a-I2-lies-in-center} suggested
to us how $I_n$ could be modified in order to get an independent
infinite sequence of conserved quantities (not Casimir invariants)
for our choice of hamiltonian; see
Sec.~\ref{s-hamiltonians-in-involution}.

\subsection{Finite Dimensional Symplectic Leaves in
$H^1(M)$}

The simplest symplectic leaf in the dual of the Lie algebra of the
rotation group is the origin of angular momentum space $\{L_i =
0\}$. It consists of a single point. Of course, one can also
characterize the orbit $\{L_i=0\}$ as the zero set of the Casimir
$L^2=0$. Can we get a similar explicit characterization of the
simplest symplectic leaves of $\svect^*$?

\subsubsection*{Orbits of Closed One Forms and Zero Set of Casimirs $I_n$}
\label{s-orbits-of-closed-forms}

The simplest symplectic leaf in $\svect^*$ should be the orbit of
exact $1$-forms. Suppose $A = d\La$ is an exact 1-form. Then under
the coadjoint action of $u \in \svect$ it goes to $A' = A + ad_u^*
A$
    \beq
        A'_i = A_i - ({\cal L}_u A)_i = \pdr_i \La - (\pdr_j \La) (\pdr_i
        u^j) + u^j \pdr_j \pdr_i \La =  \pdr_i (\La - u^j \pdr_j \La)
    \eeq
We see that a pure gauge is mapped to a pure gauge under the Lie
algebra coadjoint action. Thus the tangent space to the orbit of
exact $1$-forms is trivial. So pure gauges form a single-point
orbit.

After the pure gauges, the next simplest configurations are closed
but inexact 1-forms $dA = 0$, $A \ne d\La$. Suppose $M$ is a
manifold with non-vanishing first cohomology. What is the orbit of
an element of $H^1(M)$? In particular, is the orbit a finite
dimensional manifold? Does the orbit lie within $H^1(M)$? The
answers to both these questions is affirmative. Suppose $A$ is an
exact 1-form, $F = dA = 0$. Under the Lie algebra coadjoint action,
$A' = A + ad^*_u A = A - {\cal L}_u A$.
    \beq
        dA' = d(A- {\cal L}_u A) = - d(d i_u + i_u d) A
            = - d i_u dA = 0
    \eeq
The tangent space to the orbit of a closed 1-form contains only
closed $1$-forms. Since the space of closed 1-forms on a
two-dimensional manifold is finite dimensional, the tangent space to
the orbit must be finite dimensional. Moreover, we have shown that
the exact 1-forms form a single point orbit by themselves. Thus, at
the infinitesimal level, the orbit of a non-trivial element of
$H^1(M)$ must lie within $H^1(M) \setminus \{0\}$. This view will be
reinforced by considering the zero set of Casimirs
    \beq
        I_n = \int (dA/\mu)^n \mu = 0
    \eeq
$I_n$ are constant on coadjoint orbits. Thus, the orbits must be
contained in their level sets. $I_{2n}$ is the integral of a
positive quantity and therefore vanishes iff $dA=0$. If $dA=0$ then
$I_{2n+1}$ is also zero. Thus the level set $I_n=0$ is the space of
closed $1$-forms on $M$. Thus, the coadjoint orbit of a closed one
form must lie in $H^1(M)$. Since we already established that the
pure gauges form a single-point orbit, this means $I_n$ cannot be a
complete set of Casimirs if $H^1(M)$ is non-trivial.

For example, if $M$ is the plane, then $H^1(\mathbf{R}^2)$ consists
only of the equivalence class of pure gauge configurations and the
zero set of $I_n$ contains only one single-point orbit $[A]=0$. The
same is true of the 2-sphere $\mathbf{S}^2$ which has trivial first
cohomology. For the 2-torus $H^1(\mathbf{T}^2) \simeq \mathbf{R}^2$.
For $\mathbf{T}^2$, there must be symplectic leaves which are
submanifolds of $\mathbf{R}^2 \setminus \{0\}$. It is interesting to
find the orbits of cohomologically non-trivial gauge fields more
explicitly as well as the induced symplectic structure. However, we
do not investigate this further since the Hamiltonian we pick
(\ref{e-mag-egy-hamiltonian}) vanishes on closed 1-forms. There is
no interesting dynamics on the finite dimensional symplectic leaves
we have described above. Therefore, we turn to the case where $M$ is
simply connected and try to characterize the orbits of gauge fields
that are not closed.

\subsection{Symplectic Leaves when $M$ is Simply
Connected}

In the case of the rigid body, symplectic leaves other than $L_i =
0$ are concentric spheres of non-zero radius, all two dimensional
symplectic manifolds. These leaves may also be characterized as
non-zero level sets of $L^2$. By analogy, we would like to find the
orbits in $\svect^*$ of 1-forms that are not closed. They must lie
within non-zero level sets of $I_n$. Can we say something more about
them, such as whether they are finite dimensional? We address these
questions below, assuming $M$ is simply connected.

By Poincare's lemma, all the information in a gauge field on a
simply connected $M$ can be stored in the 2-form field strength, $F
= dA$. Then, the dual of the Lie algebra $\svect^*$ may be
identified with the space of scalar functions on $M$
    \beq
        \svect^* = \{f = \fr{dA}{\mu}\}
    \eeq
The coadjoint action of $\phi \in \sdiff$ is just the pull back
    \beq
        Ad^*_\phi f = \phi^* f = f \circ \phi
    \eeq
and the Lie algebra coadjoint action of $u \in \svect$ is
    \beq
        f \mapsto f + ad^*_u f = f -{\cal L}_u f = f - u^i \pdr_i f
    \eeq
Thus the coadjoint orbit of $f$ and the tangent space to the orbit
at $f$ are
    \beq
        {\cal O}_f &=& \{ f \circ \phi ~|~ \phi \in \sdiff \} \cr
        T_f {\cal O} &=& \{f - u^i \pdr_i f ~|~ u \in \svect \}
    \eeq
The picture that emerges from our analysis below, is that there are
two types of orbits when $M$ is simply connected. Orbits of the
first type contain only a single point, namely a constant function
$f = dA/\mu = c$. In the case of the plane with the uniform measure,
the only one-point orbit is $\{f = (dA/\mu) = 0\}$, consisting of
the pure gauge configuration. The orbits of the second type are all
infinite dimensional. They are the orbits of non-constant functions
$f$. In the case of the plane with uniform measure, we expect an
infinite number of such orbits of the second type, each contained
within a level set of $\{I_n, n = 1,2,3,\ldots\}$, though we do not
rule out the existence of more than one such orbit in any one level
set of the $I_n$.

\subsubsection*{Single Point Orbits}

We show that constant functions $f = (dA/\mu) = c~$ are the only
single point coadjoint orbits in the dual of the Lie algebra
$\svect^*$ when $M$ is simply connected. They lie within the level
sets $I_n = {\rm Vol}(M,\mu) ~ c^n $ where ${\rm Vol}(M,\mu) =
\int_M \mu$.

Suppose $f$ is a constant. Then ${\cal L}_u f = 0$. So the constant
functions form single point orbits. The case $f = 0$ corresponds to
the closed and exact 1-forms $A$ which we already identified as a
single point leaf if $M$ is simply connected. Non-zero constant
functions are not admissible elements of $\svect^*$ if
$M=\mathbf{R}^2$ and $\mu$ is the uniform measure. But they are
allowed if $M$ is compact or if $\mu \to 0$ at infinity on
non-compact $M$. Note that $I_n = \int f^n \mu = {\rm Vol}(M,\mu) ~
c^n$ if $f=c$. So the constant functions lie in the level sets with
$I_n = {\rm Vol}(M,\mu) ~ c^n$.

We can go one step further and show that constant functions are the
{\em only} single point orbits if $M$ is simply connected. Suppose
$f$ is a single point orbit. Then ${\cal L}_u f$ must vanish for all
volume preserving vector fields $u$. Since $M$ is simply connected,
any such vector field can be written in terms of a stream function
$u^i = \rho^{-1}  \veps^{ij} \pdr_j \psi$. Then denoting derivatives
by subscripts,
    \beq
        {\cal L}_u f &=& \rho^{-1}  \veps^{ij} (\pdr_j \psi) (\pdr_i f)
            = \rho^{-1} (f_x \psi_y - f_y \psi_x) \cr
        {\cal L}_u f = 0 &\Rightarrow& f_x \psi_y = \psi_x f_y
    \eeq
This must be true for all stream functions $\psi$. Taking $\psi = x$
and $\psi = y$ successively tells us that $f$ must be independent of
both $x$ and $y$ and hence a constant. {\em We conclude that the
only single point orbits in $\gd = \svect^*$ are the constant
functions $f = (dA/\mu)= c$ when $M$ is simply connected.}

\subsubsection*{Orbit and Stabilizer of Non-constant
Element of $\svect^*$.}

The stabilizer algebra ${\rm Stab}(f)$  or isotropy subalgebra of a
function $f$ is the set of all volume preserving vector fields $u^i$
which leave it fixed under the coadjoint action. The coset space
$\svect / {\rm Stab}(f)$ then has the same dimension as the tangent
space to the orbit containing $f$. For $u$ to be in ${\rm Stab}(f)$
we need ${\cal L}_u f =0$. Since $M$ is simply connected we can
express $u$ in terms of its stream function $u^i = \rho^{-1}
\veps^{ij} \pdr_j \chi$. The condition ${\cal L}_u f= 0$ becomes
    \beq
        \veps^{ij} (\pdr_i f) (\pdr_j \chi) = 0 ~~\Rightarrow~~
            \chi_x f_y - \chi_y f_x = 0
    \eeq
which in vector notation\footnote{We emphasize that this equation is
independent of any metric on $M$.} says that $\grad f \times \grad
\chi = 0$. Colloquially, the gradient of $\chi$ must be everywhere
parallel to the gradient of $f$. $\chi = c f$ is clearly a solution
for any real number $c$, so the isotropy subalgebra is at least one
dimensional, provided $f$ is not a constant. The product of two
solutions as well as real linear combinations of solutions are again
solutions to this linear PDE. To better understand the general
solution of this PDE, let us first consider the specific example of
a function $f(r,\tht)$ that is circularly symmetric.

\subsubsection*{Example: Stabilizer and Orbit of Circularly Symmetric
Function} \label{s-stab-and-orb-of-circ-symm-fn}

For example, let $f = (dA/\mu)$ be a non-constant function depending
only on the radial coordinate $r = \sqrt{x^2 + y^2}$. Let us call
its orbit by the name ${\cal O}_f$. The gradient $\grad f$ points
radially, so to speak. The isotropy subalgebra of any such function
$f$ is the space of stream functions $\chi$ with $\grad f \times
\grad \chi = 0$. The solutions are stream functions $\chi(r)$ that
are independent of $\tht$
    \beq
        {\rm Stab}(f) = \{\chi(r,\tht) ~|~ \pdr_{\tht} \chi = 0 \}
    \eeq
In this case, the isotropy subalgebra is infinite dimensional. The
coset space
    \beq
        \svect / {\rm Stab}(f) = \{\psi(r,\tht)| \psi \sim \psi
            + \chi(r) \}
    \eeq
has the same dimension as the tangent space to the orbit $T_f
{\cal{O}}_f$ at $f$. Though the stabilizer is infinite dimensional,
the orbit of $f(r)$ is infinite dimensional as well. For eg. the
coset space can be parameterized using an infinite number Fourier
coefficients
    \beq
        \psi(r,\tht) = \sum_{n=1}^\infty \psi^{(c)}_n(r) \cos{n \tht} +
        \sum_{n=1}^\infty \psi^{(s)}_n(r) \sin{n \tht}
    \eeq
We omitted the $\tht$ independent additive term in order to quotient
out by ${\rm Stab}(f)$.

To summarize, the isotropy subalgebra of a non-constant circularly
symmetric function consists of all stream functions that are
circularly symmetric. Moreover, the tangent space to the orbit may
be identified with the coset space in which two stream functions are
identified if they differ by one depending on $r$ alone.

\subsubsection*{Non-constant Functions have Infinite Dimensional Orbits}
\label{s-non-const-fns-inf-dim-orbs}

Using the circularly symmetric example as a guide, we can
characterize the isotropy subalgebra of a general non-constant
function. The condition for $\chi$ to be in the isotropy subalgebra
is $f_x \chi_y - f_y \chi_x = 0$. We will describe the set of all
solutions $\chi$ to this linear PDE in a region of $M$ where $df$ is
never the zero 2-form. Such a region is guaranteed to exist by an
analogue of the inverse function theorem, since we assumed that $f$
was not a constant. Our answer is that the isotropy subalgebra of
$f$ consists of all stream functions $\chi$ which are constant along
level sets of $f$. The level sets of $f$ are necessarily one
dimensional curves in such a region.

To see this, we argue as follows. In a region where $f(x,y)$ has no
local extrema, the level sets of $f$ are one dimensional. These
level curves foliate the region. Pick a coordinate $\Tht(x,y)$ along
the level curves. Then $\pdr_\Tht f = f_\Theta = 0$. Also pick a
coordinate $R$ `transversal' to the level curves of $f$. What this
means is that $\pdr_R$ and $\pdr_\Tht$ should be linearly
independent at each point $(R,\Tht)$. There is a lot of
arbitrariness in the choice of these coordinates, and one certainly
does not require any metric to define them. The volume form $\mu$ is
non-degenerate, so the volume enclosed by the mini-parallelogram
determined by the tangent vectors $\pdr_R$ and $\pdr_\Tht$ is
non-zero. Denoting partial derivatives by subscripts, we get
    \beq
    \mu(\pdr_R,\pdr_\Tht) &=& \rho dx \wedge dy (\pdr_R,\pdr_\Tht) \cr
    &=&   \half \rho (dx \otimes dy - dy \otimes dx) (\ov{R_x} \pdr_x
        + \ov{R_y} \pdr_y, \ov{\Tht_x} \pdr_x + \ov{\Tht_y} \pdr_y)
        \cr
    &=& \half \rho \bigg(\ov{R_x \Tht_y} - \ov{R_y \Tht_x} \bigg)
    = \half \rho \bigg(\fr{R_y \Tht_x - R_x \Tht_y}{R_x \Tht_y R_y \Tht_x }\bigg)
    \ne 0
    \eeq
Since $\rho$ is non-vanishing, this means
    \beq
        R_y \Tht_x - R_x \Tht_y \ne 0
    \eeq
The condition for $\chi$ to lie in the isotropy subalgebra becomes
(using $f_\Tht = 0$)
    \beq
    f_x \chi_y - f_y \chi_x &=& (f_R R_x + f_\Tht \Tht_x) (\chi_R R_y + \chi_\Tht
        \Tht_y) - x \leftrightarrow y \cr
    &=& f_R R_x (\chi_R R_y + \chi_\Tht
        \Tht_y) - f_R R_y (\chi_R R_x + \chi_\Tht \Tht_x) \cr
    &=& f_R \chi_\Tht (R_x \Tht_y - R_y \Tht_x)
    \eeq
We have already shown that $R_x \Tht_y - R_y \Tht_x \ne 0$.
Moreover, $f$ is non-constant and $f_\Tht =0$ so we must have $f_R
\ne 0$. We conclude that $\chi_\Tht = 0$. In other words, the stream
functions $\chi$ in the isotropy subalgebra are constant on level
sets of $f$.
    \beq
    {\rm Stab}(f) = \{ {\rm ~Stream ~functions~} \chi
        {\rm ~constant~ on~ level~ sets~ of~} f \}
    \eeq
It is clear that the space of such stream functions is closed under
linear combinations and also under multiplication, as we observed
earlier. The tangent space to the orbit at $f$ is then identified
with the coset space
    \beq
    T_f {\cal O}_f = \svect / {\rm Stab}(f) = \{\psi | \psi \sim \psi + \chi,
    ~ \chi {\rm ~constant~ on~ level~ sets ~of~} f\}
    \eeq
Both the stabilizer and the orbit are infinite dimensional, provided
$f$ is not constant, for the same reason as given in the example
where $f$ was circularly symmetric. {\em Thus, non-constant $f=
dA/\mu$ have infinite dimensional orbits.}

\section{Hamiltonian}
\label{s-hamiltonian}

Having specified observables (gauge-invariant functions), their p.b.
and phase space (coadjoint orbit of $\sdiff$), we need to pick a
gauge invariant hamiltonian.

\subsection{Hamiltonian Leading to Ideal Hydrodynamics}
\label{s-ideal-hydrodynamics}

There is more than one interesting way to pick a hamiltonian. The
classic choice leading to ideal hydrodynamics, requires a positive
metric $g_{ij}$ on $M$ which defines a positive, symmetric, inner
product on $\g = \svect$
    \beq
        \bra u,v \ket_{\g} = \int_M g(u,v) ~ \mu = \int_M g_{ij}u^i v^j ~ \mu.
    \label{e-inn-prod-hydrodyn}
    \eeq
The hamiltonian of Eulerian hydrodynamics is then
    \beq
        H(u) = \half \bra u,u \ket_{\g} ~= \half \int g_{ij} ~u^i ~u^j ~\rho~ d^2x
        \label{e-hamil-euler-hydro}
    \eeq
There is no {\it a priori} relation between the metric $g$ and the
volume form $\mu$. $\Omega_g = \sqrt{\det{g}} dx^1 \wedge dx^2$ need
not equal $\mu$. If $\mu = \Om_g$, the theory is particularly
natural as well as non-trivial. The hamiltonian defines an inertia
operator $I$ (generalization of the inertia tensor of a rigid body,
see \ref{a-dual-of-lie-alg-coadjoint-orbits},
Ref.~\cite{arnold-khesin}) from the Lie algebra $\g = \svect$ to its
dual $\gd = \Omega^1(M)/ d \Omega^0(M)$. Suppose $u,v$ are volume
preserving vector fields. Then the inertia operator is defined by
the equation $(Iu,v) = \bra u,v \ket_{\g}$. In other words,
    \beq
        \int (Iu)_j v^j \mu = \int g_{ij} u^i v^j \mu
        \Rightarrow \int ((Iu)_j - g_{ij} u^i) ~v^j ~\mu =0
    \eeq
Since this is true for an arbitrary volume preserving $v$, $((Iu)_j
- g_{ij} u^i)$ must be a total derivative: $(Iu)_j - g_{ij} u^i =
\pdr_j \La$ for some scalar function $\La$ which vanishes on the
boundary of $M$ or at infinity. We see that the metric does not
determine the 1-form $Iu$ uniquely, but rather up to an exact
1-form. Thus, the image $Iu$ of a volume preserving vector field $u
\in \g$ is an equivalence class in $\gd$ = gauge fields modulo gauge
transformations. A coset representative is given by the 1-form $A_i
= g_{ij} u^j$. The equation of motion is the well-known Euler
equation
    \beq
        \dd{u}{t} = - \grad_u u - \grad p~; ~~~~
        {\cal L}_u \mu = 0
    \eeq
Here $\grad_u u$ is the covariant derivative (with respect to
$g_{ij}$) of $u$ along itself. The two equations fix the pressure
$p(x,t)$ up to an additive constant. The Euler equations of
hydrodynamics have a geometric interpretation\cite{arnold-khesin}.
The configuration space of a volume preserving fluid flowing on the
manifold $M$ is the group of volume preserving diffeomorphisms
$\sdiff$. The inner product (\ref{e-inn-prod-hydrodyn}) on the Lie
algebra of this group can be extended to a right-invariant metric on
the whole group by right translations by group elements. Then, by
the least action principle, the time evolution of the fluid is given
by geodesics on $\sdiff$ with respect to this metric.

\subsection{Hamiltonian as Magnetic Energy}
\label{s-hamiltonian-as-mag-egy}

Now we propose a hamiltonian different from that of Eulerian
hydrodynamics. It is a magnetic energy, natural from the point of
view of Yang-Mills theory. The result will still be a theory of
geodesics on $\sdiff$, but with respect to a different
right-invariant metric on this group than that implied by ideal
hydrodynamics. Suppose $M$ is a 2-dimensional surface with volume
form $\mu$ and metric $g_{ij}$. The metric did not play any role so
far since the phase space and poisson structure are independent of
it. But to specify the dynamics, we need the metric. Any two
dimensional manifold is conformally flat, so the information in
$g_{ij}$ is encoded in its volume form $\Omega_g = \sqrt{g} dx^1
\wedge dx^2$, where $g = \det{g_{ij}}$. We {\em do not} assume that
$\mu$ is equal to $\Omega_g$. Indeed, if that is the case, the
dynamics is trivial. By analogy with Yang-Mills theory, we pick the
manifestly gauge-invariant magnetic energy
    \beq
    H = \half \int \bigg(\fr{F \wedge *F}{\Om_g }\bigg) \mu
    \eeq
as our hamiltonian. It can be written in a variety of equivalent
ways
    \beq
    H &=& \half \int (F / \Omega_g)^2 \mu
        = \half \int (*F)^2 \mu
        = \half \int (F/\mu)^2 \sigma \mu \cr
        &=& \ov{4}  \int F_{ij} F^{ij} \mu
        = \half \int \bigg(\fr{B^2 \sigma}{\rho}\bigg) d^2x.
    \label{e-mag-egy-hamiltonian}
    \eeq
We find the last formula most useful in calculations. Here $\sigma$
is the positive scalar function on $M$ given by the square of the
quotient of the two volume forms.
    \beq
    \sigma = (\mu / \Omega_g)^2 = \rho^2/g
    \eeq
We will see that for the dynamics to be non-trivial, $\sigma$ must
not be constant. The hamiltonian in (\ref{e-mag-egy-hamiltonian}) is
the simplest choice that is gauge-invariant, quadratic in gauge
fields and involves two derivatives. It is easy to see the
equivalence of the various formulae for the magnetic energy in
(\ref{e-mag-egy-hamiltonian}). Let $\eps^{ij} = g^{ik} g^{jk}
\eps_{kl}$,  $\eps^{ij} = - \eps^{ji}$, $\eps^{12} = 1/g$. Moreover,
$g \eps^{ij} = \veps^{ij}$. The Hodge dual of $F$ is the scalar
    \beq
        *F &=& *(dx^i \wedge dx^j) \half F_{ij} = \half \sqrt{g}
            \eps^{ij} F_{ij} = \ov{2 \sqrt{g}} \sum_{ij} F_{ij}
            \eps_{ij} = \fr{B}{\sqrt{g}}
    \eeq
$F \wedge *F $ is the 2-form $(B^2 / \sqrt{g}) dx^1 \wedge dx^2$.
$((F \wedge *F)/ \Om_g) = {B^2 /g}$ is a scalar. Hence
    \beq
        (F \wedge *F) / \Om_g = (*F)^2 = (F / \Om_g)^2 = \fr{B^2}{g} =
        \fr{B^2 \sig}{\rho^2} = (F/\mu)^2 \sig
    \eeq
Moreover, $F_{ij} F^{ij} = 2 (F_{12} F^{12})$. But $F^{12} = g^{1i}
g^{2j} F_{ij} = (g^{11} g^{22} - g^{12} g^{21}) F_{12} = g^{-1}
F_{12} = B/g$. Thus $F_{ij} F^{ij} = 2 {B^2 / g}$. This shows the
equivalence of all the forms of the hamiltonian given in
(\ref{e-mag-egy-hamiltonian}).

The hamiltonian (\ref{e-mag-egy-hamiltonian}) involves only the
spatial components of the field strength since there is no time
component for the gauge field in our theory. So there is no analogue
of electric energy. Moreover, our theory is not relativistically
covariant unlike electrodynamics. The dynamics determined by
(\ref{e-mag-egy-hamiltonian}) is {\em not} equivalent to Eulerian
hydrodynamics, though both theories share the same phase space. To
see this, note that if $\Omega_g = \mu$, then
(\ref{e-mag-egy-hamiltonian}) reduces to the Casimir $I_2$ and has
trivial dynamics while the hamiltonian (\ref{e-hamil-euler-hydro})
continues to have non-trivial dynamics.

The differential of the hamiltonian (\ref{e-mag-egy-hamiltonian}) is
($\dd{B(y)}{A_i(x)} = - \veps^{ij} \pdr_j \delta^2(x-y)$)
    \beq
    (dH(A))^i =  \ov{\rho} \fr{\delta H}{\delta A_i(x)}
            = \ov{\rho} \veps^{ij} \pdr_j(B \sigma/\rho),
    \eeq
For each $A$, $dH(A)$ is volume preserving $\pdr_i(\rho dH^i) =
\veps^{ij} \pdr_i \pdr_j (B\sig/\rho) = 0$. $H$ generates a
1-parameter (time) family of diffeomorphisms of a coadjoint orbit of
$\sdiff$, which serves as the phase space. The hamiltonian vector
field $V_H$ is a vector field on a coadjoint orbit $\cal O$. At each
tangent space $T_{[A(x)]} {\cal O}$ to an orbit, $V_H([A])$ is given
by the coadjoint action of the Lie algebra element $dH(A)$,  $V_H =
ad^*_{dH(A)}$. The change in an observable $f$ under such an
infinitesimal canonical transformation is the Lie derivative with
respect to $V_H$
    \beq
    \fr{d f(A)}{dt} = -{\cal L}_{V_H} f = ad_{dH}^* f = \{H,f\} = (A,[dH,df]) = \int A_i [dH,df]^i ~\mu
    \eeq
Explicit equations of motion in local coordinates are given in
Sec.~\ref{s-eqn-motion}.

\subsection{Inertia Operator and Inner Product on $\gd$ from
Hamiltonian}

Recall (Sec.~\ref{s-ideal-hydrodynamics}) that the hamiltonian of
ideal hydrodynamics (\ref{e-hamil-euler-hydro}) defines an inner
product on the Lie algebra of volume preserving vector fields via
the inertia operator. Here, the magnetic energy defines a positive
inner product $\bra.,. \ket_{\gd}$ on the dual of the Lie algebra,
$\g^* = \Omega^1(M) / d \Omega^0(M)$, via an inverse inertia
operator $h$, obtained below. This inner product is degenerate if
$M$ has non-vanishing first cohomology. Away from degeneracies, one
can in principle invert $h$ to obtain an inertia operator and an
inner product $\bra.,. \ket_{\g}$ on the Lie algebra $\g = \svect$.
It should be possible to extended $\bra.,. \ket_{\g}$ to a metric on
the group $G = \sdiff$ by means of right translations. The equations
of motion in $\gd = \svect^*$ determined by $H$ should be the
projection of the geodesic flow on $G= \sdiff$ with respect to this
right-invariant metric. In this sense, the dynamics determined by
the magnetic energy describes geodesics on $\sdiff$. Eulerian
hydrodynamics also describes geodesics on $\sdiff$, but with respect
to a different right-invariant metric\cite{arnold-khesin}. Here, we
obtain the inverse of the inertia operator and the inner product on
$\svect^*$ explicitly. Let us begin by writing the hamiltonian as a
quadratic form on gauge fields. The result is
    \beq
        H = \half \int A_i ~h^{il}~ A_l ~\rho ~d^2x {\rm ~~with~~}
        h^{il} = \fr{\veps^{ij} \veps^{kl}}{\rho} \bigg[ (\pdr_j
        \fr{\sigma}{\rho}) \pdr_k + \fr{\sigma}{\rho} \pdr_j \pdr_k
        \bigg]
    \label{e-inv-inertia-op}
    \eeq
$h^{ij}$ plays the role of inverse inertia operator, mapping
equivalence classes of gauge fields to volume preserving vector
fields by raising the index $A_j \mapsto h^{ij} A_j = (dH(A))^i$. To
get (\ref{e-inv-inertia-op}), note that $H$ may be written in terms
of its differential,
    \beq
        H = \half (A, dH(A)) = \half \mu_{dH}(A) = \half \int A_i ~(dH(A))^i~ \rho ~
        d^2x.
    \eeq
This is easily checked by integrating by parts assuming $B \to 0$ on
$\pdr M$:
    \beq
        H &=& \half \int (B^2 \sigma / \rho) d^2x = \fr{\veps^{ij}}{2} \int (B \sigma /
        \rho) \pdr_i A_j d^2x = \fr{\veps^{ij}}{2} \int
        \pdr_j(B \sigma / \rho) A_i d^2x \cr &=& \half \int A_i dH^i
        \rho d^2x
    \eeq
As a consequence,
    \beq
        H = \half \int A_i (h A)^i \mu
        {\rm ~~where~~}
        (hA)^i = h^{ij} A_j = dH^i = \fr{\veps^{ij}}{\rho}
            \pdr_j(B \sigma / \rho)
    \eeq
Writing this out we get,
    \beq
        (hA)^i = \fr{\veps^{ij} \veps^{kl}}{
            \rho} \pdr_j\bigg(\fr{\sigma}{\rho} \pdr_k A_l \bigg)
            = \fr{\veps^{ij} \veps^{kl}}{\rho} \bigg[ (\pdr_j (\sigma / \rho))
            \pdr_k + (\sigma / \rho) \pdr_j \pdr_k \bigg] A_l
    \eeq
From this we read off the inverse of the inertia operator
(\ref{e-inv-inertia-op}). Through $h$, the magnetic energy naturally
defines a symmetric positive inner product on the dual of the Lie
algebra, $\g^* = \Omega^1(M) / d \Omega^0(M)$. This inner product
may be written in the following equivalent ways
    \beq
        \bra A,\tilde A \ket_{\g^*} = (A, h \tilde A) = \mu_{h\tilde A}(A) =
        \int A_i h^{ij} {\tilde A}_j ~ \rho ~d^2x = \int B~ \tilde
        B~ \bigg( \fr{\sigma}{\rho}\bigg) d^2x
    \label{e-inner-prod}
    \eeq
Here $B,\tilde B$ are the magnetic fields corresponding to $A,
\tilde A$. Positivity of the inner product is ensured since
$\sigma(x) = ({\mu / \Omega_g})^2 = (\rho^2/g) \geq 0$ and the
integrand is positive
    \beq
        \bra A, A \ket_{\g^*} = 2 H(A) = \int (B^2 \sigma / \rho) d^2x \geq 0
    \eeq
Symmetry of the inner product is shown by establishing the last
equality in (\ref{e-inner-prod}):
    \beq
        \int A_i (hA)^i \rho d^2x &=&  \veps^{ij} \int A_i \pdr_j(\tilde B \sigma /
        \rho) d^2x = -  \veps^{ij} \int \pdr_j A_i (\tilde B \sigma /
        \rho) d^2x \cr && = \int (B \tilde B \sigma / \rho) d^2x.
    \eeq
Inner product (\ref{e-inner-prod}) is degenerate precisely on gauge
fields in the first cohomology of $M$, $H^1(M) = \{ B= 0, A \ne d\La
\}$. If $M$ is simply connected, (\ref{e-inner-prod}) is a
non-degenerate inner product on $\gd = \svect^*$. Inverting it gives
an inner product on $\g$ which may be extended by right translation
to a right invariant metric on $\sdiff$.

\section{Equations of Motion}
\label{s-eqn-motion}

Before working out the equations of motion in detail, it would be
prudent to convince ourselves that unlike the Casimir $I_2$, the
magnetic energy $H$ does {\em not} lie in the center of the Poisson
algebra for non-constant $\sigma$. We show this in
\ref{a-H-not-in-center} by exhibiting a moment map that has
non-vanishing p.b. with $H$.

\subsection{Time Evolution of Magnetic Field}

The time evolution of the magnetic field is given by the equation
    \beq
    \dot{B} = \{H,B\} =  \veps^{ij} \pdr_i(B/\rho) \pdr_j(B\sig/\rho)
        = \grad (B/\rho) \times \grad(B\sig/\rho)
    \label{e-time-ev-of-mag-fld}
    \eeq
where $\sig = {\rho^2 / g}$. It is manifestly gauge-invariant.
Though the hamiltonian is quadratic, (\ref{e-time-ev-of-mag-fld}) is
non-linear since the p.b. of gauge fields (\ref{e-pb-of-gauge-flds})
is linear in gauge fields. The `interactions' are partly encoded in
the Poisson structure and partly in the hamiltonian, so to speak.
(\ref{e-time-ev-of-mag-fld}) is an analogue of the Euler equation
$\dot{L} = L \times \Om,~ L = I \Om$ for the rigid body, (see
\ref{a-dual-of-lie-alg-coadjoint-orbits}). Indeed, the equation of
motion for the magnetic field can be regarded as the Euler equation
for the group $\sdiff$ with respect to the metric
(\ref{e-inner-prod}) defined by the hamiltonian
(\ref{e-mag-egy-hamiltonian}). The time evolution of the magnetic
field can also be written as an equation of motion for the field
strength $F = dA$ defined in (\ref{e-defn-of-fld-strength-mag-fld})
    \beq
        \dot{F} = \half  \veps^{ij} \pdr_i(F/\mu) \pdr_j(F \sig/\mu)
        ~\eps_{kl} dx^k \wedge dx^l
    \eeq
The formula (\ref{e-time-ev-of-mag-fld}) for $\dot{B}$ is obtained
using the p.b. formula (\ref{e-pb-formula}) using the relation
$\fr{\delta B(z)}{\delta A_j(y)} = - \veps^{jk}
\dd{\delta^2(z-y)}{y^k}$, assuming $A = 0$ on $\pdr M$. After some
integration by parts,
    \beq
        \fr{d B}{dt} = \{H,B\} = - \veps^{il} \veps^{jk} \pdr_k \bigg[
            \rho^{-1} (\pdr_i A_j) + A_i (\pdr_j \rho^{-1})
            + \ov{\rho} A_i \pdr_j \bigg] \pdr_l (B \sigma /\rho)
    \eeq
Expanding out derivatives of products, eliminating $A$ in favor of
$B = \veps^{ij} \pdr_i A_j$ and after a lot of cancelations, one
arrives at the advertised evolution equation.

\subsection{Time Evolution of Moment Maps}

We have already argued that the Casimirs $I_n = \int (dA/\mu)^n
~\mu$ are constant on symplectic leaves. Thus, they are conserved
under time evolution $\fr{d I_n}{dt} = \{H,I_n\} = 0$, independent
of the choice of hamiltonian. As for the moment maps $\mu_u$, we can
show using (\ref{e-pb-formula}) that
    \beq
        \fr{d \mu_u(A)}{dt} = \{H,\mu_u\} = \int d^2x A_i
            \bigg[ \veps^{jk} (\pdr_j u^i) +  \veps^{ik} u^j \rho^{-1}
            (\pdr_j \rho) -  \veps^{ik} u^j \pdr_j
            \bigg] \pdr_k(B \sigma / \rho)
    \label{e-time-ev-of-mom-maps}
    \eeq
For $\rho =1$ this reduces to
    \beq
        \fr{d \mu_u(A)}{dt} = \int d^2x A_i
            \bigg[ \veps^{jk} (\pdr_j u^i) -  \veps^{ik} u^j \pdr_j
            \bigg] \pdr_k(B \sigma)
    \eeq
The rhs is gauge-invariant even though the gauge field appears
explicitly.

\subsection{Infinitely Many Conserved Charges in Involution}
\label{s-hamiltonians-in-involution}

We find an infinite set of conserved charges
    \beq
     \fr{d H_n}{dt} =  \{H,H_n\} = 0; ~~~ n = 1,2,3,\ldots
    \eeq
for a uniform $\mu$ ($\rho = 1$) and $g_{ij}$ an arbitrary
metric\footnote{$\sigma = \rho^2/g$ must not grow too fast at
$\infty$ and the magnetic field must vanish at $\infty$.}. We
suspect that a similar result holds for non-uniform $\mu$. The
conserved quantities are
    \beq
    H_n = \int (dA/\mu)^n ~\sigma ~\mu = \int B^n ~\sigma ~d^2x,
    \eeq
where $\sig = (\mu / \Om_g)^2 = (1/g)$ and $g = \det{g_{ij}}$.
Furthermore, we find that $H_n$ are in involution $\{H_m,H_n\} = 0$.
The hamiltonian is $H = \half H_2$. The presence of infinitely many
conserved charges in involution suggests an integrable structure
underlying the dynamics determined by the magnetic energy. The
hamiltonian of ideal hydrodynamics (\ref{e-hamil-euler-hydro}) is
not known nor expected to have any conserved quantities besides the
$I_n$ and their close relatives, which are constant on coadjoint
orbits. What is remarkable about $H_n$ is that they are {\em not}
constant on coadjoint orbits but still conserved quantities for the
magnetic energy hamiltonian (\ref{e-mag-egy-hamiltonian}). To
establish this we investigate the time evolution of $H_n$.
    \beq
    \fr{d H_n}{dt} = \{H,H_n\} = \int A_i \bigg[ \fr{\delta H}{\delta
        A_j} \pdr_j \bigg( \rho^{-1} \fr{\delta H_n}{\delta
        A_i}\bigg) - \fr{\delta H_n}{\delta A_j} \pdr_j \bigg(
        \rho^{-1} \fr{\delta H}{\delta A_i}\bigg)
        \bigg] d^2x
    \eeq
Using
    \beq
    \fr{\delta H_n}{\delta A_i} = n  \veps^{ij} \pdr_j \bigg(
        \sigma (B/\rho)^{n-1} \bigg);  ~~~~~~~
    \fr{\delta H}{\delta A_i} =  \veps^{ij} \pdr_j \bigg(
        \sigma B/\rho \bigg)  ~~~~~~~
    \eeq
We get
    \beq
    \fr{d H_n}{dt} &=& n \veps^{jk} \veps^{il} \int A_i
        \bigg[\pdr_k(\sigma B / \rho) \pdr_j(\rho^{-1} \pdr_l(\sigma (B/\rho)^{n-1}))
        \cr && - \pdr_k(\sigma (B/\rho)^{n-1}) \pdr_j (\rho^{-1} \pdr_l (\sigma B / \rho)) \bigg] d^2x
    \eeq
For $\rho = 1$ this becomes
    \beq
    \fr{d H_n}{dt} = n \veps^{jk} \veps^{il} \int A_i
        \bigg[\pdr_k(\sigma B) \pdr_j \pdr_l(\sigma B^{n-1})
        - \pdr_k(\sigma B^{n-1}) \pdr_j \pdr_l (\sigma B ) \bigg] d^2x
    \eeq
In the gauge $A_1 = 0$ , $A_2 \equiv A$, $B = \pdr_x A$ this may be
written as
    \beq
    \fr{d H_n}{dt} &=& n \int dx dy A \bigg[ \pdr_x(B\sigma) \pdr_{xy} (B^{n-1}
        \sigma) - \pdr_x(B^{n-1} \sigma) \pdr_{xy}(B \sigma) \cr &&-
        \pdr_y(B\sigma) \pdr_x^2(B^{n-1} \sigma) + \pdr_y(B^{n-1}
        \sigma) \pdr_x^2 (B\sigma)
        \bigg]
    \label{e-time-ev-of-Hn}
    \eeq
We prove in \ref{a-involution} that $H_n, ~n = 1, 2, 3, \ldots$ are
conserved by showing that the rhs of (\ref{e-time-ev-of-Hn})
vanishes. Having found an infinite number of conserved quantities we
wanted to know whether their p.b. generates new conserved
quantities. In our case, we discovered (again for $\rho = 1$), that
the conserved quantities $H_n$ are in involution, i.e. they mutually
Poisson commute $ \{H_m,H_n\} = 0$. The proof of this is somewhat
lengthy and is relegated to \ref{a-involution}.

Intuitively, $H_n$ are independent of each other since they are like
average values of different powers of $B$. Furthermore, $H_n$ are
independent of the Casimirs $I_m$.  $I_m$ contain no information
about the metric $g_{ij}$ while $H_n$ depend on the metric via
$\sigma$. For example, we show in \ref{a-H-not-in-center} that $H_2$
is not a Casimir. More generally, it would be nice to prove that
$I_n$ and $H_m$ are functionally independent by showing that on
every tangent space to an orbit, the cross product of their
gradients is non-vanishing.

\section{Static Solutions}
\label{s-static-solns}

\subsection{Zero Energy Configurations}

The magnetic energy is $H = \int_M (dA/\mu)^2 \sigma \mu = \int (B^2
/\rho) \sigma d^2x$ where $\sigma = (\mu / \Omega_g)^2 = (\rho^2 /
g) \geq 0$. Thus, the energy is non-negative, $H \geq 0$. Moreover,
if $A$ is closed, the energy automatically is a global minimum $dA =
0 \Rightarrow H = 0$. Moreover, since $H$ is the integral of the
square of $dA$, weighted by a positive function, closed gauge fields
are the only configurations with zero energy. Any such closed gauge
field is a static solution to the equations of motion (irrespective
of $\mu$ and $g_{ij}$)
    \beq
    \fr{df}{dt}|_{dA=0} = \{H,f\}|_{dA=0} = 0
    \eeq
As discussed in Sec.~\ref{s-orbits-of-closed-forms}, the closed
gauge fields constitute one or more (according as $M$ is simply
connected or not) symplectic leaves of the Poisson manifold. They
correspond to the zero set of Casimirs $I_n = \int (dA/\mu)^n \mu =
0$. Thus, the hamiltonian vanishes on all the symplectic leaves with
$I_n=0$ and there is no interesting dynamics to speak of. If $M$ is
the plane, then the pure gauges are the only ones with zero energy.
The leaf/leaves with $I_n=0$ are the analogue of the $L^2 = 0$ point
at the origin $L_i =0$ of the angular momentum Poisson manifold. At
that point, the energy $E = \sum_i L_i^2/ 2 I_i$ of the rigid body
vanishes as well.

It is interesting to find minima of energy on more interesting
symplectic leaves. Initial conditions determine which symplectic
leaf is the phase space of the theory. The general problem of
finding the minimum of energy on a given symplectic leaf (perhaps
specified through values of invariants such as $I_n$) is potentially
quite interesting and difficult. One would first have to find which
gauge fields or magnetic field configurations satisfy the
constraints and lie on the specified leaf. This is similar to the
problem we solved in the large $N$ limit of 2d QCD where we found
the minimum of energy on the symplectic leaf with baryon number
equal to one\cite{gsk-thesis,soliton-parton}. Here we do the
opposite, find a few static solutions and then determine which orbit
they lie on.

\subsection{Circularly Symmetric Static Solutions}

Suppose $M = \mathbf{R}^2$. If both the volume forms $\mu$ and
$\Omega_g$ are circularly symmetric, then any circularly symmetric
initial magnetic field will remain unchanged with time. To see this,
note that $\sigma = \rho^2/g$ depends only on the radial coordinate
$r$. Suppose that at $t=0$, $B(r,t=0)$ depends only on $r$. The
initial value problem for $B$ given in (\ref{e-time-ev-of-mag-fld})
is
    \beq
        \dot{B} = \grad (B/\rho) \times \grad(B\sig/\rho)
    \eeq
Due to circular symmetry, both the gradients point radially and
their cross product vanishes. Thus $\dot{B}=0$ and we have a static
solution $B(r)$!

We already met the $B(r)=0$ static solution before, it lies on a one
point symplectic leaf of pure gauge configurations. However, the
static solutions corresponding to non-constant $B(r)$ lie on
infinite dimensional symplectic leaves which we found previously
(Sec.~\ref{s-stab-and-orb-of-circ-symm-fn}). The values of Casimirs
on the orbit containing a circularly symmetric static solution
$B(r)$ are
    \beq
        I_n = 2\pi \int_0^\infty (B(r)/\rho(r))^n ~\rho(r) ~r ~dr
    \eeq
By a judicious choice\footnote{Finding $B(r)$ for given $I_n,
\rho(r)$ is similar to the Classical Moment Problem.} of $B(r)$ one
should be able to find a static solution $B(r)$ for given $\rho(r)$
that lies on an orbit with practically any desired value for the
invariants $I_n$. More generally, by an argument similar to the one
given in Sec.~\ref{s-non-const-fns-inf-dim-orbs}, we see that
magnetic fields for which $(B/\rho)$ and $(B \sig/ \rho)$ have
common 1-dimensional level sets, are static solutions of
(\ref{e-time-ev-of-mag-fld}).

\subsection{Some Other Local Extrema of Energy}

For a gauge field to be an extremum of energy on a given symplectic
leaf, the variation of energy in directions tangential to the leaf
must vanish. There is no need for variations in other, let alone
all, directions to vanish. However, though it is not necessary, if
$[A]$ is such that all variations of the hamiltonian $H(A)$ vanish,
then, $A$ must be a local extremum of energy. Such an extremum
$\fr{\delta H}{\delta A_k} = 0$ is automatically a static solution
to the equations of motion for any gauge-invariant observable $f$
    \beq
        \fr{d f}{dt}  = \{H, f\} = \int d^2x d^2y~
            \{A_i(x),A_j(y)\} \fr{\delta H}{\delta A_i(x)}
            \fr{\delta f}{\delta A_j(y)} = 0
    \eeq
These extrema are given by solutions of
    \beq
        \fr{\delta H}{\delta A_k} = 2  \veps^{ij} \pdr_j(B \sigma / \rho)
            = 0 ~~\Rightarrow ~~\pdr_1 (B \sigma /\rho) = 0 {\rm ~and~} \pdr_2 (B \sigma /\rho) = 0
    \eeq
The only solutions are $B = (c \rho / \sig) = (c g / \rho)$ where
$c$ is a constant and $g = \det{g_{ij}}$. These extrema lie on
leaves where the Casimirs take the values
    \beq
    I_n ~=~ c^n \int \bigg(\fr{\Om_g}{\mu }\bigg)^{2n} \mu ~=~ c^n \int
    \sig^{-n}~ \rho~ d^2x
    \eeq
We do not yet understand the physical meaning of these extrema of
energy. They have a finite energy if $B = (c \rho/\sigma)$ vanishes
at infinity sufficiently fast.

\section{Discussion}
\label{s-discussion}

A summary of the paper was given in the introduction
Sec.~\ref{s-intro}. Here, we mention a few directions for further
study. We would like to know whether there is a deeper integrable
structure that would explain the presence of an infinite number of
conserved charges in involution for the uniform volume measure
$\mu$. The extension to an arbitrary volume form seems likely. Are
there any time-dependent exact solutions of the non-linear evolution
equation? What is the Poisson algebra of loop observables and can
the Hamiltonian be written in terms of them? Can this model be
quantized and is there a non-abelian extension? An extension to
$3+1$ dimensions is possible, though the Poisson algebra has only
one analytically known Casimir, the Hopf or link invariant. Is there
a Lorentz covariant theory along these lines? Can the idea that
gauge fields be thought of as dual to volume preserving vector
fields be exploited in any other context?

How is our gauge theory related to hydrodynamics and turbulence?
Recently, Jackiw, {\it et. al.}\cite{Jackiw:2004nm} have studied
perfect fluids and certain non-abelian extensions. Our gauge theory
shares the same phase space as ideal hydrodynamics, but the two
theories have different hamiltonians. However, similar methods may
be useful in the study of both theories. For example, Iyer and
Rajeev\cite{2d-turbulence-rmt} (see also Sec.~11.D of
Ref.~\cite{arnold-khesin}) have proposed a statistical approach to
$2$ dimensional turbulence, based on a matrix regularization of the
phase space. It may be possible to use a similar regularization for
our gauge theory. For an $N\times N$ matrix regularization to be
integrable, it would appear that we need ${\cal O}(N^2)$ conserved
quantities, while $I_n$ and $H_m$ furnish only $2N$ conserved
quantities. It is unclear what this implies for the integrability of
the continuum theory we have proposed in this paper. On the other
hand, Polyakov has proposed a theory of turbulence in $2+1$
dimensions based on conformal
invariance\cite{polyakov-conf-turbulence}. Since the group of
conformal transformations and area preserving transformations are
disjoint except for isometries, it appears unlikely that there is
any direct relation of our work to Polyakov's.

\section*{Acknowledgements}

This investigation was motivated by discussions with S. G. Rajeev.
The author also benefitted from discussions with J. Ambjorn, G.
Arutyunov, G. 't Hooft and J. Noldus and thanks G. Arutyunov, G. 't
Hooft, L. F. Alday and S. G. Rajeev for comments on the manuscript.

\appendix

\section{Poisson Manifolds and Coadjoint Orbits}
\label{a-dual-of-lie-alg-coadjoint-orbits}

We collect a few facts about classical
mechanics\cite{arnold-class-mech,arnold-khesin} that we use, to make
the paper self-contained and fix notation.

\subsection{Poisson Manifolds and Symplectic Leaves}

The basic playground of classical mechanics is a Poisson manifold.
It is a manifold $M$ with a product $\{.,.\} ~:~ {\cal F}(M) \times
{\cal F}(M) \longrightarrow {\cal F}(M)$ (the Poisson bracket p.b.)
on the algebra of observables. $\{.,.\}$ is bilinear, skew symmetric
and satisfies the Jacobi identity and Leibnitz rule. ${\cal F}(M)$
is a class of real-valued functions (say, $C^\infty(M)$). Any such
function $f : M \to {\mathbf{R}}$ generates canonical
transformations on $M$. A canonical transformation is a flow on $M$,
associated to the canonical vector field $V_f$. The Lie derivative
of any function $g(A)$ along the flow is given by the p.b. ${\cal
L}_{V_f} g(A) = \{f,g\}(A), ~A \in M.$ Flow lines of the canonical
transformation generated by $f$ are integral curves of $V_f$.

Often, Poisson algebras of observables are degenerate. They have a
center (Casimirs) which have zero p.b. with {\em all} observables.
In such a situation, the Poisson manifold as a whole cannot serve as
the phase space of a physical system, since it would not be a
symplectic manifold. Rather, it is the symplectic leaves of a
Poisson manifold that can serve as phase spaces. The symplectic leaf
of a point $A \in M$ is the set of all points of $M$ reachable from
$A$ along integral curves of canonical vector fields. On a
symplectic leaf, the Poisson structure is non-degenerate and can be
inverted to define a symplectic structure, a non-degenerate closed
2-form $\om$. Indeed, if $\xi,\eta$ are tangent vectors at $A$ to a
symplectic leaf, then the symplectic form at $A$ is $\om(\xi,\eta) =
\{f,g\}(A)$ where $f$ and $g$ are {\em any} two functions whose
canonical vector fields at $A$ coincide with $\xi$ and $\eta$; i.e.
$\xi = V_f |_A, \eta = V_g |_A$. Moreover, on any symplectic leaf,
$\om(V_f,.) = df(.)$ where $df$ is the exterior derivative of $f$.

The hamiltonian vector field $V_H$ is the canonical vector field of
the hamiltonian $H: M \to {\mathbf{R}}$. $V_H$ generates time
evolution $\fr{df}{dt} = \{H,f\} =  {\cal L}_{V_H} f$. Irrespective
of the hamiltonian, time evolution always stays on the same
symplectic leaf.

\subsection{Coadjoint Orbits in the Dual of a Lie Algebra}
\label{a-coadj-orb-as-symp-lvs-in-gdual}

A natural example of a Poisson manifold, that occurs in many areas
of physics, is the dual of a Lie algebra. The symplectic leaves of
the dual of a Lie algebra are the coadjoint orbits of the group. To
understand this, suppose $G$ is a group, $\g$ its Lie algebra and
$\gd$ the dual of the Lie algebra. Then we have a bilinear pairing
between dual spaces $(A,u) \in {\mathbf{R}}$ for $A \in \gd$ and $u
\in \g$. Suppose $f,g$ are two real-valued functions on $\gd$. Then
their p.b. is defined using the differential $df(A) \in \g$
    \beq
    \{f,g\}(A) = (A,[df,dg])
    \eeq
$[df,dg]$ is the commutator in $\g$. This turns $\gd$ into a Poisson
manifold, which is often degenerate. The symplectic leaves are
coadjoint orbits of the action of $G$ on $\gd$. To see this, first
define the inner automorphism $A_g : G \to G, ~~ A_g h = g h g^{-1}$
which takes the group identity $e$ to itself. The group adjoint
representation $Ad_g : \g \to \g$ is the linearization of the inner
automorphism at $e$: $Ad_g = {A_g}_*|_e$ and is $Ad_g u = g u
g^{-1}$ for matrix groups. The group coadjoint representation
$Ad_g^* : \gd \to \gd$ is defined as $(Ad_g^* A, u) = (A, Ad_g u)$.
The group coadjoint orbit of $A \in \gd$ is
    \beq
    {\cal O}_A = \{ Ad^*_g A ~| ~ g \in G \}
    \eeq
The Lie algebra adjoint representation is $ad_u : \g \to \g$ where
$ad_u = \fr{d}{dt}|_{t=0} Ad_{g(t)}$ for a curve $g(t)$ on the group
with $g(0) = e$ and $\dot{g}(0) = u$ and takes the form $ad_u v =
[u,v]$ for matrix groups. The Lie algebra coadjoint representation
$ad^*_u: \gd \to \gd$  is defined by $(ad_u^* A,v) = (A,ad_u v)$.
The Lie algebra coadjoint orbit of $A \in \gd$ is the tangent space
at $A$ to the group coadjoint orbit of $A$.
    \beq
    \{ ad_u A ~|~ u \in \g \} = T_A {\cal O}_A
    \eeq
Thus, a tangent vector $\xi$ to a coadjoint orbit at $A$ may be
written as $\xi = ad^*_u A$ for some (not necessarily unique) $u \in
\g$.

There is a natural symplectic structure on coadjoint orbits, which
turns them into homogeneous symplectic leaves of $\gd$. The
symplectic form (Kirillov form) $\omega$ acting on a pair of tangent
vectors to the orbit ${\cal O}_A$ at $A$ is given by $\om(\xi,\eta)
= (A,[u,v])$ where $u,v \in \g$ are {\em any} two Lie algebra
elements such that $\xi = ad^*_u A$ and $\eta = ad^*_v A$. Thus, the
coadjoint orbits are symplectic manifolds which foliate $\gd$ in
such a way as to recover the Poisson structure on the whole of
$\gd$. The different coadjoint orbits in $\gd$ are not necessarily
of the same dimension, but are always even dimensional if their
dimension is finite.

The canonical vector field $V_f$ of an observable $f : \gd \to
\mathbf{R}$ at a point $A \in \gd$ is given by the Lie algebra
coadjoint action of the differential $df(A)$, $V_f(A) = ad^*_{df} A$
    \beq
    {\cal L}_{V_f} g(A) = \{f,g\}(A) =
    (A,[df,dg]) = (A, ad_{df} dg) = (ad^*_{df} A , dg)
    \eeq
In particular, infinitesimal time evolution is just the coadjoint
action of the Lie algebra element $dH$, the differential of the
hamiltonian. For example, if $\mu_u(A) = (A,u)$ is the {\em moment
map} for $u \in \g$ , then the canonical vector field $V_{\mu_u}(A)
= ad_u^* A$. In other words, infinitesimal canonical transformations
generated by moment maps are the same as Lie algebra coadjoint
actions.

Observables in the center of the Poisson algebra (Casimirs) are
constant on coadjoint orbits. They are invariant under the group and
Lie algebra coadjoint actions. To show that an observable is a
Casimir, it suffices to check that it commutes with the moment maps
which generate the coadjoint action.

\noindent {\bf Example: Eulerian Rigid Body} Let $\Om^i$ be the
components of angular velocity of a rigid body in the co-rotating
frame. The components of angular velocity lie in the Lie algebra
$\g$ of the rotation group $G = SO(3)$. The dual space to angular
velocities consists of angular momenta $L_i$ with the pairing (or
moment map) $(L,\Om) = L_i \Om^i$. The space of angular momenta is
the dual $SO(3)^* = \mathbf{R}^3$. The latter carries a Poisson
structure $\{L_i, L_j \} = \eps_{ijk} L_k$. Observables are
real-valued functions of angular momentum $f(L)$ and satisfy the
p.b.
    \beq
    \{f, g \}(L) = \sum_{i,j} \{L_i,L_j\} \dd{f}{L_i} \dd{g}{L_j}
    \eeq
The space of angular momenta $\mathbf{R}^3$ must be a degenerate
Poisson manifold, since it is not even dimensional. Indeed, the
symplectic leaves are concentric spheres centered at the point $L_i
= 0$ as well as the point $L_i = 0$. These symplectic leaves are the
coadjoint orbits of $SO(3)$ acting on the space $\mathbf{R}^3$ of
angular momenta. The symplectic form on a sphere of radius $r$ is
given by $r \sin \tht d\tht \wedge d\phi$. The Casimirs are
functions of $L^2 = \sum_i L_i^2$ and are constant on the symplectic
leaves, they are invariant under the coadjoint action of the
rotation group on $\mathbf{R}^3$. The hamiltonian is $H = \sum_i
L^2_i/2 I_i$ if the axes are chosen along the principle axes of
inertia. $I_i$ are the principle moments of inertia, the eigenvalues
of the inertia operator $I_{ij}: \g \to \gd$ which maps angular
velocities to angular momenta $L_i = I_{ij} \Om^j$. The equations of
motion $\dot{L} = ad^*_{dH} L  = ad^*_{I^{-1}L} L = ad^*_{\Om} L = L
\times \Om$ are $\dot{L_i} = \{H,L_i\}$, $\dot{L_1} = a_{23} L_2
L_3$ and cyclic permutations thereof, where $a_{ij} = I_j^{-1} -
I_i^{-1}$.

\section{The Charges $I_n$ are in Involution}
\label{a-In-are-in-involution}

We show that the charges $I_n$ are in involution $\{ I_m, I_n \} =
0$. To calculate
    \beq
        \{ I_m, I_n \} = \int_M A_i \bigg[\fr{\delta I_m}{\delta A_j} \pdr_j \bigg(\ov{\rho}
            \fr{\delta I_n}{ \delta A_i} \bigg)  - \fr{\delta I_n}{\delta A_j} \pdr_j \bigg(\ov{\rho}
            \fr{\delta I_m}{\delta A_i} \bigg) \bigg] d^2x
    \eeq
we need
    \beq
        I_m = \int (B/\rho)^m \rho d^2x ~~\Rightarrow~~
        \fr{\delta I_m}{\delta A_i} = m  \veps^{ij} \pdr_j ((B/\rho)^{m-1})
    \eeq
So the p.b. becomes
    \beq
        \{ I_m, I_n \} = mn \veps^{il} \veps^{jk} \int_M d^2x ~ A_i \bigg[
        \pdr_k (B / \rho)^{m-1} \pdr_j \bigg(\ov{\rho} \pdr_l (B / \rho)^{n-1} \bigg)
            -  m \leftrightarrow n \bigg]
    \eeq
The first term in square brackets $\pdr_k (B / \rho)^{m-1} \pdr_j
\bigg(\ov{\rho} \pdr_l (B / \rho)^{n-1} \bigg)$ can be written as
    \beq
        && (m-1)(n-1) (B/\rho)^{m-2} \pdr_k(B/\rho) \pdr_j \bigg\{
            \rho^{-1} (B/\rho)^{n-2} \pdr_l(B/\rho)
            \bigg\} \cr
        &=& (m-1)(n-1) (B/\rho)^{m+n-4} \pdr_j (\rho^{-1})
            \pdr_k(B/\rho) \pdr_l(B/\rho) \cr
        && + ~ (m-1)(n-1)(n-2) (B/\rho)^{m+n-5} \rho^{-1} \pdr_k(B/\rho)
            \pdr_j(B/\rho) \pdr_l(B/\rho) \cr
        && + ~ (m-1)(n-1)(B/\rho)^{m+n-4} \rho^{-1} \pdr_k(B/\rho)
        \pdr_j \pdr_l(B/\rho)
    \eeq
The $1^{\rm st}$ and $3^{\rm rd}$ terms are symmetric under $m
\leftrightarrow n$ and therefore do not contribute to $\{ I_m, I_n
\}$. Therefore we get ($I_1 = 0$, so we can ignore $m,n = 1$)
    \beq
        \fr{\{ I_m, I_n \}}{mn(m-1)(n-1) (n-m)} = \int d^2x
        \fr{(B/\rho)^{m+n-5}}{\rho} \bigg[ \veps^{il} \veps^{jk}
        A_i \pdr_l(B/\rho) \pdr_j(B/\rho)
        \pdr_k(B/\rho) \bigg]
    \nonumber
    \eeq
Now the term in square brackets vanishes identically due to
antisymmetry of $ \veps^{jk}$. We conclude that $\{ I_m, I_n \} =
0$. Thus, $I_n, n = 1,2,3 \ldots$ are an infinite number of charges
in involution.

\section{$I_2$ is a Casimir of the Poisson Algebra}
\label{a-I2-lies-in-center}

To find the infinitesimal change of $I_n$ under the coadjoint
action, we calculate $\{I_n,\mu_u\}$. If this vanishes for all
volume preserving $u$, then $I_n$ would be constant on symplectic
leaves and hence a Casimir. Recall (\ref{e-differential-of-In}) that
the differential of $I_n$ is
    \beq
        (dI_n(A))^i = n \rho^{-1}  \veps^{ij} \pdr_j(B/\rho)^{n-1}
    \eeq
and the differential of the moment map is $(d \mu_u)^i = u^i$. Their
Lie bracket is
    \beq
        [u,dI_n]^i &=& u^j \pdr_j(n \rho^{-1}  \veps^{ik}
            \pdr_k (B/\rho)^{n-1}) - n \rho^{-1}  \veps^{jk}
            \pdr_k (B/\rho)^{n-1} \pdr_j u^i \cr
        &=& n \bigg[  \veps^{ik} u^j \pdr_j (\rho^{-1}
            \pdr_k (B/\rho)^{n-1}) - \rho^{-1}  \veps^{jk}
            \pdr_k (B/\rho)^{n-1} \pdr_j u^i \bigg]
    \eeq
Thus $\{\mu_u,I_n\} = \int A_i  [u,dI_n]^i \rho d^2 x$ gives
    \beq
        \{\mu_u,I_n\}
        = n \int d^2x A_i \bigg[  \veps^{ik} \rho u^j \pdr_j (\rho^{-1}
        \pdr_k (B/\rho)^{n-1})  -  \veps^{jk} \pdr_k (B/\rho)^{n-1}
        \pdr_j u^i \bigg].
    \label{e-pb-of-In-and-mom-map}
    \eeq
For $\rho = 1$ this becomes
    \beq
        \{\mu_u,I_n\} = n \int d^2x A_i \bigg[  \veps^{ik} u^j (\pdr_k
        \pdr_j  B^{n-1}) -  \veps^{jk} (\pdr_j u^i) (\pdr_k B^{n-1}) \bigg]
    \eeq
Specializing to $n=2$, and writing $u$ in terms of its stream
function, $u^i =  \veps^{il} \pdr_l \psi$,
    \beq
        \{I_2,\mu_u\} = 2\int d^2x \bigg[
        \veps^{il} \veps^{jk} A_i (\pdr_l \pdr_j \psi)(\pdr_k B)
        - \veps^{ik} \veps^{jl} A_i (\pdr_k \pdr_j B) (\pdr_l \psi)
            \bigg]
    \eeq
$\{I_2,\mu_u\}$ is gauge-invariant, so we calculate it in the gauge
$A_1 = 0$, $B = \pdr_1 A_2$ and denote $A_2 = A$, $x^1 = x$, $x^2 =
y$ and derivatives by subscripts.
    \beq
        \half \{I_2,\mu_u\} = \int dx dy A \bigg[\psi_y A_{xxx} - \psi_x A_{xxy}
                + \psi_{xy} A_{xx} - \psi_{xx} A_{xy} \bigg]
    \eeq
The idea is to integrate by parts and show that this expression
vanishes. Let us temporarily call the second factor of $A$ by the
name $\A$. Write $\half \{I_2,\mu_u\}$ as a sum of four terms $T_1 +
T_2 + T_3 + T_4$
    \beq
        T_1 = \int dx dy A \psi_y \A_{xxx}; &&
        T_2 = - \int dx dy A \psi_x \A_{xxy} \cr
        T_3 = \int dx dy A \psi_{xy} \A_{xx}; &&
        T_4 = - \int dx dy A \psi_{xx} \A_{xy}.
    \eeq
We will show that $T_1 + T_2 = - T_3 - T_4$. Integrating by parts
till there are no derivatives on $\A$,
    \beq
        T_1 &=& -\int dx dy \A \bigg(\psi_y A_{xxx} + 3 A_{xx} \psi_{xy}
            + 3 A_x \psi_{xxy} + A \psi_{xxxy} \bigg) \cr
        T_2 &=& \int dx dy \A \bigg( \psi_x A_{xxy} + 2 A_{xy} \psi_{xx}
            + 2 A_x \psi_{xxy} + A_{xx} \psi_{xy} + A_y \psi_{xxx} + A \psi_{xxxy}
            \bigg) \cr
        T_3 &=& \int dx dy \A \bigg( \psi_{xy} A_{xx} + 2 A_x \psi_{xxy}
            + A \psi_{xxxy} \bigg) \cr
        T_4 &=& -\int dx dy \A \bigg( A_{xy} \psi_{xx} + A_y
        \psi_{xxx} + A_x \psi_{xxy} + A \psi_{xxxy} \bigg)
    \eeq
Using the fact that $\A = A$ we get for $T_1$  and $T_2$
    \beq
        2 \times T_1 &=& - \int dx dy \A \bigg( 3 A_{xx} \psi_{xy}
            + 3 A_x \psi_{xxy} + A \psi_{xxxy} \bigg) \cr
        2 \times T_2 &=& \int dx dy \A \bigg(2 A_{xy} \psi_{xx}
            + 2 A_x \psi_{xxy} + A_{xx} \psi_{xy} + A_y \psi_{xxx} + A \psi_{xxxy}
            \bigg)
    \eeq
$T_3$ and $T_4$ give us the identities
    \beq
        \int dx dy ~ \A A_x \psi_{xxy} &=& - \int dx dy~ \A A
            \psi_{xxxy} \cr
        \int dx dy ~ \A A_y \psi_{xxx} &=& \int dx dy ~ \A A_x \psi_{xxy}
    \eeq
Use these to simplify $T_1$ and $T_2$ by eliminating $\psi_{xxxy}$
and $\psi_{xxx}$ in favor of $\psi_{xxy}$:
    \beq
        2 \times T_1 &=& - \int dx dy ~ \A \bigg( 3 A_{xx} \psi_{xy}
            + A_x \psi_{xxy} \bigg) \cr
        2 \times T_2 &=& \int dx dy \A \bigg( 2 A_{xy} \psi_{xx}
            + A_{xx} \psi_{xy} + A_{x} \psi_{xxy} \bigg)
    \eeq
Adding these and setting $A = \A$
    \beq
        T_1 + T_2 = \int dx dy ~ A \bigg( A_{xy} \psi_{xx}
            - A_{xx} \psi_{xy} \bigg)
    \eeq
Meanwhile by definition,
    \beq
        T_3 + T_4 &=& \int dx dy ~ A \bigg( A_{xx} \psi_{xy} -
            A_{xy} \psi_{xx}  \bigg) \cr
    {\rm Thus~~ } \{I_2,\mu_u\} &=& 2 (T_1 + T_2 + T_3 + T_4)  = 0
    \eeq
{\em We conclude that $I_2$ lies in the center of the Poisson
algebra for constant $\rho$.}

\section{$H = \half \int (dA/\mu)^2 \sigma \mu$ is {\em not} a Casimir}
\label{a-H-not-in-center}

$I_2 = \int (dA /\mu)^2 \mu$ turned out to be in the center of the
Poisson algebra of gauge-invariant functions. Here we show that the
magnetic energy $H = \half \int (dA/\mu)^2 \sigma \mu$ with a
non-constant $\sig = \rho^2/g$, {\em does not} lie in the center,
and therefore leads to non-trivial time evolution. We do this by
giving an explicit example of a gauge-invariant function with which
it has a non-vanishing p.b. Consider $\rho =1$, then from
(\ref{e-time-ev-of-mom-maps})
    \beq
        \{H,\mu_u\} = \int d^2x A_i \bigg[  \veps^{jk} (\pdr_j u^i)
        \pdr_k(B\sigma) -  \veps^{ik} u^j (\pdr_j \pdr_k B \sigma)
        \bigg]
    \eeq
In gauge $A_1= 0$, $A_2 \equiv A$ and with $u^i = \veps^{ij} \pdr_j
\psi$ we get ($x^1 = x, x^2 = y$)
    \beq
        \{H,\mu_u\} &=& \int dx dy A \bigg[ -(\pdr_x^2 \psi) \pdr_y (B
        \sigma) + (\pdr_{xy} \psi) \pdr_x (B\sigma) \cr && + (\pdr_y \psi)
        \pdr_x^2(B \sigma) - (\pdr_x \psi) \pdr_{xy} (B\sigma)
        \bigg]
    \eeq
Now for the simple choices $\psi = xy$, $A = \sigma =
e^{-(x^2+y^2)/2}$ we have $B = \pdr_x A = -x e^{-(x^2 + y^2)/2}$.
The p.b. can be calculated exactly to yield
    \beq
        \{H,\mu_u\} = \int dx dy e^{-3(x^2 + y^2)/2} \bigg(-1 - 4x^4 - 2y^2
            + 4x^2(2+y^2) \bigg) = \fr{2\pi}{27}
    \eeq
{\em Thus $H$ does not lie in the center of the Poisson algebra.} We
also checked that $H$ transforms non-trivially under many other
generators $\mu_u$ of the coadjoint action.

\section{$H_n$ are in Involution for Uniform Measure}
\label{a-involution}

Suppose $\mu$ is the uniform measure ($\rho =1$). We prove that the
charges
    \beq
        H_n = \int (dA/\mu)^n \sigma \mu  = \int (B/\rho)^n \sigma
            \rho d^2x  = \int B^n \sigma d^2x
    \eeq
are in involution
    \beq
        \{H_m,H_n\} = 0 {\rm ~~for~~} m,n = 0,1,2,3, \ldots
    \eeq
An immediate corollary is that $H_n$ are conserved quantities since
the hamiltonian is $\half H_2$. Here $B \to 0$ at $\infty$ and
$\sigma = (\mu/\Om_g)^2 = 1/g$ must be such that these integrals
converge. The proof involves explicitly computing the p.b. and
integrating by parts several times. Using
    \beq
    \fr{\delta H_n}{\delta A_i}= \delta^i H_n = n  \veps^{ij} \pdr_j(\sigma (B/\rho)^{n-1})
    \eeq
we can express the p.b. as
    \beq
    \{H_m,H_n\} &=& \int A_i \bigg[ \delta^j H_m \pdr_j(\rho^{-1} \delta^i
        H_n) - m \leftrightarrow n \bigg] d^2x
    = mn \veps^{il} \veps^{jk} \int A_i \cr &\times&  \bigg[ \pdr_k(\sigma
        (B/\rho)^{m-1}) \pdr_j(\rho^{-1} \pdr_l(\sigma (B/\rho)^{n-1}))
        - m\leftrightarrow n   \bigg] d^2x
    \eeq
For $\rho =1$ this becomes
    \beq
    \fr{\{H_{m+1},H_{n+1}\}}{(m+1)(n+1)} = \veps^{il} \veps^{jk} \int A_i
    \bigg[\pdr_k(\sigma B^m) \pdr_j \pdr_l (\sigma B^n) - m\leftrightarrow n   \bigg] d^2x
    \eeq
Now we expand out the derivatives of products of $B$ and $\sig$ and
eliminate terms that vanish due to antisymmetry of $ \veps^{jk}$ or
antisymmetry in $m$ and $n$. We get
    \beq
    \fr{\{H_{m},H_{n}\}}{mn(n-m)} &=& \veps^{il} \veps^{jk}  \int A_i
        B^{m+n-4} \bigg[B (\pdr_k \sig) (\pdr_l \sig) (\pdr_j B)
        + \sig B (\pdr_k \sig) (\pdr_j \pdr_l B)
    \cr && + (n+m-3) \sig (\pdr_k
        \sig) (\pdr_j B) (\pdr_l B) + \sig B (\pdr_j B)(\pdr_k \pdr_l \sig)
        \bigg]
    \eeq
Since this p.b. is gauge-invariant, calculate in the gauge $A_1 =
0$, call $A_2 = A$, $B = \pdr_1 A$ and denote derivatives by
subscripts ($x^1 = x, x^2 = y$)
    \beq
    \{H_{m+1},H_{n+1}\} &=& -  \veps^{jk} (m+1) (n+1)(n-m) \int dxdy~ A
        B^{m+n-2} \bigg[ B \sig_x B_j \sig_k
    \cr && + \sig B B_{jx} \sig_k
        + (n+m-1) \sig B_x B_j \sig_k + \sig B B_j \sig_{kx}
        \bigg].
    \eeq
After collecting terms, this becomes
    \beq
    \{H_{m},H_{n}\} &=& mn(m-n) \int A B^{m+n-4} \bigg[ (B_x \sig_y - B_y
        \sig_x)  (B \sig_x + (n+m-3) B_x \sig) \cr && + \sig B \bigg\{ B_{xx}
        \sig_{y} - B_{xy} \sig_x + B_x \sig_{xy} - B_y \sig_{xx}
        \bigg\} \bigg].
    \eeq
To simplify it we define $k = m+n -4$ and use identities such as
$B^k B_x = (B^{k+1})_x / (k+1)$ and $\sigma \sigma_x = \half
(\sigma^2)_x$ to combine the factors of $B$ and $\sigma$ to write
    \beq
    \ov{mn (m-n)} \{H_{m},H_{n}\} &=& \sum_{i=1}^8 T_i ~~~~ {\rm where} \cr
    T_1 = \ov{k+2} \int A (B^{k+2})_x \sig_x \sig_y &&
    T_2 = -\ov{k+2} \int A (B^{k+2})_y \sig_x \sig_x \cr
    T_3 = \half \int A (B^{k+1})_x B_x (\sig^2)_y  &&
    T_4 = - \half \int A (B^{k+1})_y B_x (\sig^2)_x \cr
    T_5 = \half \int A B^{k+1} B_{xx} (\sig^2)_y  &&
    T_6 = -\half \int A B^{k+1} B_{xy} (\sig^2)_x \cr
    T_7 = \ov{k+2} \int A (B^{k+2})_x \sig \sig_{xy} &&
    T_8 = -\ov{k+2} \int A (B^{k+2})_y \sig \sig_{xx}
    \eeq
Numerically we find $T_1 +  T_7 + T_2 + T_8 = -(T_3 + T_5 + T_4 +
T_6)$. To see this analytically, integrate by parts with the aim of
eliminating $A$ in favor of $B = \pdr_x A$. Using similar identities
as before,
    \beq
        T_1 &=& - T_7 - \half \int \fr{(B^{k+3})_x (\sig^2)_y}{k+3}
            - \half \int \fr{A (B^{k+2})_{xx} (\sig^2)_y}{k+2} \cr
        T_2 &=& - T_8 + \half \int \fr{(B^{k+3})_y (\sig^2)_x}{k+3}
            + \half \int \fr{A (B^{k+2})_{xy} (\sig^2)_x}{k+2} \cr
        T_3 &=& - T_5 - \half \int \fr{(B^{k+3})_x (\sig^2)_y}{k+3}
            - \half \int \fr{A (B^{k+2})_x (\sig^2)_{xy}}{k+2} \cr
        T_4 &=& - T_6 + \half \int \fr{A_y (B^{k+2})_x (\sig^2)_x}{k+2}
            + \half \int \fr{A (B^{k+2})_x (\sig^2)_{xy}}{k+2}
    \eeq
Then
    \beq
        T_3 + T_5 + T_4 + T_6 &=& - \half \int \fr{(B^{k+3})_x
        (\sig^2)_y}{k+3} + \half \int \fr{A_y (B^{k+2})_x (\sig^2)_x}{k+2 } \cr
        T_1 + T_7 + T_2 + T_8 &=& \ov{2(k+2)} \int A \bigg\{
            (B^{k+2})_{xy} (\sig^2)_x - (B^{k+2})_{xx} (\sig^2)_y
            \bigg\} \cr
        \{H_m,H_n\} &=& mn (m-n) (U_1 + U_2 + U_3 + U_4)
    \eeq
where $T_3 + T_5 + T_4 + T_6   \equiv U_1 + U_2 $ and $T_1 + T_7 +
T_2 + T_8 \equiv U_3 + U_4$. Finally, integration by parts shows
that
    \beq
        U_2 = - U_3 - \ov{2 (k+2)} \int A (B^{k+2})_x (\sig^2)_{xy}
    \eeq
and that $U_2 + U_3 = - U_4 - U_1$. We conclude that $\sum_{i=1}^4
U_i = 0$ and therefore $\{H_m,H_n\} = 0$. Since the hamiltonian is
half the second charge, $H = \half H_2$, we have shown that $H_n$
for $n=1,2,3, \ldots$ are an infinite number of conserved
quantities, which moreover, are in involution!



\begin{thebibliography}{10}


\bibitem{mandelstam}
  S.~Mandelstam,
``Feynman Rules For Electromagnetic And Yang-Mills Fields From The
Gauge Independent Field Theoretic Formalism,''
  Phys.\ Rev.\  {\bf 175} (1968) 1580.


\bibitem{rajeev-ictp}
S.~G.~Rajeev, ``Two-dimensional meson theory,'' UR-1252, Proc. 1991
Summer School in High Energy Physics and Cosmology, ed. E. Gava {\it
et. al.} World Scientific (1992) p. 955.


\bibitem{rajeev-turgut}
S.~G.~Rajeev and O.~T.~Turgut, ``Poisson algebra of Wilson loops in
four-dimensional Yang-Mills theory,'' Int.\ J.\ Mod.\ Phys.\ A {\bf
10}, 2479 (1995)  [arXiv:hep-th/9410053].



\bibitem{karabali-nair-kim}
  D.~Karabali, C.~j.~Kim and V.~P.~Nair,
``On the vacuum wave function and string tension of Yang-Mills
theories  in (2+1) dimensions,''
  Phys.\ Lett.\ B {\bf 434} (1998) 103
  [arXiv:hep-th/9804132].



\bibitem{lee-rajeev-closed}
C.~W.~H.~Lee and S.~G.~Rajeev, ``A Lie algebra for closed strings,
spin chains and gauge theories,''
  J.\ Math.\ Phys.\  {\bf 39} (1998) 5199
  [arXiv:hep-th/9806002]; 


\bibitem{lee-rajeev-open}
C.~W.~H.~Lee and S.~G.~Rajeev, ``Symmetry algebras of large-N matrix
models for open strings,''
  Nucl.\ Phys.\ B {\bf 529} (1998) 656
  [arXiv:hep-th/9712090].



\bibitem{tHooft-large-N}
G.~'t Hooft, ``A Planar Diagram Theory For Strong Interactions,''
  Nucl.\ Phys.\ B {\bf 72} (1974) 461.



\bibitem{2dqhd}
S.~G.~Rajeev, ``Quantum hadrondynamics in two-dimensions,''
  Int.\ J.\ Mod.\ Phys.\ A {\bf 9}, 5583 (1994)
  [arXiv:hep-th/9401115].



\bibitem{tHooft-2dmesons}
G.~'t Hooft, ``A Two-Dimensional Model For Mesons,''
  Nucl.\ Phys.\ B {\bf 75} (1974) 461.



\bibitem{soliton-parton}
V.~John, G.~S.~Krishnaswami and S.~G.~Rajeev, ``Parton Model from
Bi-local Solitonic Picture of the Baryon in two-dimensions,''
  Phys.\ Lett.\ B {\bf 492} (2000) 63
  [arXiv:hep-th/0310014].


\bibitem{gsk-thesis}
G.~S.~Krishnaswami, ``Large-N limit as a classical limit: Baryon in
two-dimensional QCD and multi-matrix models,'', Ph.D. Thesis,
University of Rochester (2004);
  arXiv:hep-th/0409279.



\bibitem{rajeev-turgut-der-free-alg}
S.~G.~Rajeev and O.~T.~Turgut, ``Poisson brackets of Wilson loops
and derivations of free algebras,''
  J.\ Math.\ Phys.\  {\bf 37} (1996) 637
  [arXiv:hep-th/9508103].



\bibitem{entropy-var-ppl}
L.~Akant, G.~S.~Krishnaswami and S.~G.~Rajeev, ``Entropy of
operator-valued random variables: A variational principle  for large
N matrix models,''
  Int.\ J.\ Mod.\ Phys.\ A {\bf 17} (2002) 2413
  [arXiv:hep-th/0111263];


\bibitem{coll-potn-hamiltonian-mat-mod}
A.~Agarwal, L.~Akant, G.~S.~Krishnaswami and S.~G.~Rajeev,
``Collective potential for large-N Hamiltonian matrix models and
free  Fisher information,''
  Int.\ J.\ Mod.\ Phys.\ A {\bf 18} (2003) 917
  [arXiv:hep-th/0207200].






\bibitem{arnold-khesin}
V. I. Arnold and B. A. Khesin, {\em Topological Methods in
Hydrodynamics}, Applied Mathematical Sciences 125, Springer (1998);

\bibitem{khesin}
B. A. Khesin, ``Topological fluid dynamics'',  Notices Amer. Math.
Soc. 52 (2005),  no. 1, 9--19.


\bibitem{kdv-as-euler} V.Yu. Ovsienko, and B. A. Khesin, ``Korteweg-de Vries
super-equation as an Euler equation'', Funct. Anal. Appl. {\bf 21}
(1987), no. 4, 329-331.


\bibitem{arnold-inf-dim-groups-hydrodyn} V. I. Arnold,
``Sur la g\'{e}om\'{e}trie diff\'{e}rentielle des groupes de Lie de
dimension infinie et ses applications \`{a} l'hydrodynamique des
fluides parfaits'', Ann. Inst. Fourier (Grenoble) {\bf 16} (1966) 1,
319-361


\bibitem{arnold-hamilton-euler-hydro} V. I. Arnold, ``The
Hamiltonian nature of the Euler equation in the dynamics of the
rigid body and of an ideal fluid,'' Uspekhi Mat. Nauk {\bf 24}
(1969), no. 3, 225-226;



\bibitem{novikov} S. P. Novikov, ``The Hamiltonian formalism and a many-valued
analogue of Morse theory'', Rus. Math. Surveys {\bf 37} (1982), no.
5, 1-56.


\bibitem{m-w} J. E. Marsden and A. Weinstein, ``Coadjoint orbits,
vortices, and Clebsch variables for incompressible fluids'', Physica
D 7 (1983) no. 1-3, 305-323.

\bibitem{d-k-n} B. A. Dubrovin, I. M. Krichever and S. P. Novikov,
Encyclopaedia of Math. Sci. {\bf 4} (1990), Springer Verlag,
173-280.

\bibitem{marsden} D. D. Holm, J. E. Marsden and T. S. Ratiu,
``Euler-Poincar\'{e} Models of Ideal Fluids with Nonlinear
Dispersion'' Phys. Rev. Lett., 349, (1998) 4173-4177.





\bibitem{unrelated-but-similar-sounding-1}
E.~I.~Guendelman, E.~Nissimov and S.~Pacheva, ``Volume preserving
diffeomorphisms versus local gauge symmetry,'' Phys.\ Lett.\ B {\bf
360} (1995) 57  [arXiv:hep-th/9505128];


\bibitem{unrelated-but-similar-sounding-2}
  M.~Kreuzer,
 ``Gauge Theory Of Volume Preserving Diffeomorphisms,''
  Class.\ Quant.\ Grav.\  {\bf 7} (1990) 1303.


\bibitem{kirillov} A. A. Kirillov, ``Unitary representations of
nilpotent Lie groups'', Rus. Math. Surv. {\bf 17} (1962), no. 4,
57-101.


\bibitem{weinstein-poisson} A. Weinstein, ``The local structure of
Poisson manifolds'', J. Diff. Geom. {\bf 18} (1983), no. 3, 523-557;
J. Diff.Geometry {\bf 22} (1985), no. 2, 255.



\bibitem{Jackiw:2004nm}
  R.~Jackiw, V.~P.~Nair, S.~Y.~Pi and A.~P.~Polychronakos,
``Perfect fluid theory and its extensions,''
  J.\ Phys.\ A {\bf 37} (2004) R327
  [arXiv:hep-ph/0407101].



\bibitem{2d-turbulence-rmt}
  S.~V.~Iyer and S.~G.~Rajeev,
``A Model of Two Dimensional Turbulence Using Random Matrix
Theory,''
  Mod.\ Phys.\ Lett.\ A {\bf 17} (2002) 1539
  [arXiv:physics/0206083].


\bibitem{polyakov-conf-turbulence}
  A.~M.~Polyakov,
``The Theory of turbulence in two-dimensions,''
  Nucl.\ Phys.\ B {\bf 396} (1993) 367
  [arXiv:hep-th/9212145].


\bibitem{arnold-class-mech}
V. I. Arnold {\em Mathematical Methods of Classical Mechanics}, 2nd
ed. Graduate Texts in Mathematics, Springer (1989).




\end{thebibliography}
\end{document}